\documentclass[journal]{IEEEtran}
\IEEEoverridecommandlockouts

\usepackage{amsmath,amsfonts,amsthm}
\usepackage{dsfont}
\usepackage{cite}
\usepackage[vlined,linesnumbered,ruled]{algorithm2e}
\usepackage{graphicx}
\graphicspath{ {./figures/} }

\theoremstyle{plain}

\DeclareMathOperator*{\E}{\mathds{E}}
\DeclareMathOperator*{\prob}{\mathds{P}}

\DeclareMathOperator*{\eqdef}{\stackrel{\triangle}{=}}

\def\BibTeX{{\rm B\kern-.05em{\sc i\kern-.025em b}\kern-.08em
    T\kern-.1667em\lower.7ex\hbox{E}\kern-.125emX}}

\begin{document}

\title{ Data-driven Bandwidth Adaptation \\ for Radio Access Network Slices
\thanks{.}
}

\author{\IEEEauthorblockN{Panagiotis Nikolaidis and Asim Zoulkarni and John Baras\\}
\IEEEauthorblockA{Department of Electrical \& Computer Engineering and the Institute for Systems Research\\
University of Maryland, College Park, MD 20742, USA\\
Email: \{nikolaid, asimz, baras\}@umd.edu}}

\maketitle

\begin{abstract}
The need to satisfy the QoS requirements of multiple network slices deployed at the same base station poses a major challenge to network operators. The problem becomes even harder when the desired QoS involves packet delays. In that case, network utility maximization is not directly applicable since the utilities of the slices are unknown. As a result, most related works learn online the utilities of all slices and how to split the resources among them. Unfortunately, this approach does not scale well for many slices. Instead, it is needed to perform learning separately for each slice. To this end, we develop a bandwidth demand estimator; a network function that periodically receives as input the traffic of the slice and outputs the amount of bandwidth that its MAC scheduler needs to deliver the desired QoS. We develop the bandwidth demand estimator for QoS involving packet delay metrics based on a model-based reinforcement learning algorithm. We implement the algorithm on a cellular testbed and conduct experiments with time-varying traffic loads. Results show that the algorithm delivers the desired QoS but with significantly less bandwidth than non-adaptive approaches and other baseline online learning algorithms.
\end{abstract}

\section{Introduction}
\label{intro}
A key aspect of networking slicing is the formation of a Service Level Agreement (SLA) between the Network Operator (NO) and the tenant requesting the Network Slice (NS). The SLA includes the Quality of Service (QoS) guarantees that the NO needs to provide and the cost of the NS that the tenant needs to pay. This paradigm differentiates network slicing from traditional networking, e.g., the Internet, since the former involves QoS guarantees while the latter is best-effort. 

The main mechanism that enables QoS guarantees is resource provisioning. In the  Radio Access Network (RAN) part of a cellular network, the NO needs to provision Physical Resource Blocks (PRBs) at the Base Station (BS) to ensure that the Medium Access Control (MAC) schedulers of the NSs can deliver the desired QoS. 

A simple provisioning approach is to allocate a fixed and exclusive amount of PRBs to each NS. Unfortunately such an approach is quite wasteful. Indeed, it leads to overprovisioning since the traffic of a NS varies over time and thus its MAC scheduler does not always need the same amount of PRBs to deliver the desired QoS.

Given that licensed spectrum is a scarce resource that NOs obtain through auctions, the previous approach may not be feasible for a large number of NSs. A more efficient approach is to dynamically adapt the PRBs at the MAC scheduler of each NS based on its current traffic. The operators may then exploit the time-varying PRB allocations in the NSs to reduce the overall provisioned PRBs via statistical multiplexing.

To this end, most works in the related literature devise schemes that split the resources between NSs based on network utility maximization. However, for packet delay requirements, the utility function that relates the allocated bandwidth to the packet delay metric of interest is unknown. As a result, the schemes must jointly learn online the network utilities and split the resources optimally. Notice that such approaches do not scale well as the number of NSs increases since the state-action space of the learning problem explodes.

Also, due to the presence of explicit QoS requirements, network utility maximization is not sufficient in network slicing. It is also needed to provision a sufficient amount of total bandwidth at the BS. Indeed, the fact that a customer receives their fair share of resources based on utility maximization is of little value to them if their share is not large enough to deliver the desired QoS. Unfortunately, it is not clear how to perform resource provisioning for schemes that learn online how to maximize the network utility. Notice that if the learned optimal splitting did not manage to satisfy all the SLAs due to under-provisioning, then the whole learning process must be repeated until the correct amount of bandwidth is provisioned. 

Due to the above issues, we argue that a new system architecture needs to be introduced at the BSs which is composed by two network functions; the Bandwidth Demand Estimator (BDE) and the Network Slice Multiplexer (NSM). The BDE periodically monitors the traffic state of the NS and then estimates its bandwidth demand; the amount of PRBs currently needed at its MAC scheduler to deliver the desired QoS. The NSM receives all the bandwidth demands and decides which ones to accept given the limited bandwidth at the BS and the SLAs of the NSs.

The proposed architecture enables learning to happen separately for each NS and thus is much more scalable. Once the statistics of the bandwidth demands are known, resource provisioning is greatly simplified. For instance, we may provision bandwidth that equals the $P^H$-percentile of the sum of the bandwidth demands where $P^H$ is close to $100$. We note that a NSM and a resource provisioning mechanism have been proposed already in \cite{isolation}. Here, we only focus on the design of a BDE that performs dynamic bandwidth adaptations which is currently missing in the related literature.

Specifically, we develop a BDE that estimates the required bandwidth when the desired QoS is to bound some statistic of each user's packet delays by a constant. This constant is defined in the SLA of the NS. The development of such a BDE is challenging due to difficulties in modeling the traffic of a NS, its radio channel conditions, and in analyzing the complicated service disciplines used by the MAC schedulers.

We note that there are no readily available queueing theoretic models that we may leverage to estimate the packet delays in such settings. Thus, we resort to data-driven online learning methods. The BDE proposed here is based on Reinforcement Learning (RL). To see why RL is a suitable approach, notice that small bandwidth allocations may result in large queues at the MAC scheduler which indicates that larger bandwidths will be needed in the future to satisfy the desired QoS. Thus, current actions affect the future and the problem can be naturally formulated as a Dynamic Programming (DP) problem. However, the system dynamics are not known a priori and thus a RL approach is needed.

In our RL algorithm, the state of the system is the traffic state of the NS composed by the number of active users, the average of their Modulation and Coding Schemes (MCSs) and the number of bits awaiting transmission at the MAC scheduler. The actions correspond to PRB amounts. The cost is the number of allocated PRBs plus a large constant which is multiplied by a binary variable that equals $1$ if the QoS is violated. The large constant handles the tradeoff between allocating a small bandwidth and satisfying the QoS. We provide insight on how this constant can be selected.

We emphasize that a major consideration in the design of the RL algorithm is that it needs to learn a good policy quickly. Otherwise, the users may disconnect from the NS due to long intervals of poor QoS. Thus, it is crucial to incorporate any domain knowledge in the logic of the RL algorithm to enhance its performance. Here, we consider two simple insights.

First, we assume that higher bandwidth allocations are expected to improve the QoS since MAC schedulers used in practice are work conserving. To leverage this insight, we consider a model-based RL approach where the transition matrix of the system is estimated periodically. At the end of each period, value iteration is performed to obtain the optimal policy. Then, we use the monotonicity of the cost w.r.t. the allocated bandwidth to estimate the transition probabilities of state-action pairs not seen so far. This allows us to speed up the estimation process of the transition matrix.

Second, we consider that small packet delays imply small queue lengths due to Little's Law. Thus, state-action pairs that produce low immediate costs may also have high $Q$ values. For this reason, we initialize our RL algorithm with a variation of the Upper Confidence Bound 1 (UCB1) that also exploits the monotonicity of the QoS metric w.r.t. the allocated bandwidth. As a result, we quickly obtain multiple samples for the transition probabilities of promising state-action pairs which further enhances the performance of the RL algorithm.

We implement and test the proposed algorithm on a 3GPP compliant LTE testbed composed by the AMARI Callbox and the AMARI UE Simbox \cite{callbox,uesimbox}. The AMARI Callbox is an integrated PC that implements the LTE BS and Core Network (CN). The AMARI UE Simbox is another integrated PC that emulates multiple UEs. It includes traffic and radio channel simulators. The two PCs communicate with each other over-the-air via LTE using Software Defined Radios (SDRs). Thus, the testbed can be viewed as a small scale implementation of an LTE network. Although the Amarisoft testbed can also emulate a 5G network, it does not provide enough configuration options to implement our BDE in a 5G system. Hence, we can only consider an LTE system. However, our algorithm may also be used in 5G testbeds that are more amenable to configuration than our closed-source testbed.

In summary, our contribution is a BDE that estimates periodically the bandwidth needed at the MAC scheduler of a NS to deliver the desired QoS, enabling online bandwidth adaptation in RANs. As mentioned before, the BDE is one of the two network functions that is needed to enable the proposed system architecture which provides a scalable way to satisfy the SLAs of multiple NSs with non-trivial QoS requirements. To the best of our knowledge, we are the first to develop a BDE for NSs in the RAN and showcase its performance on a testbed that is 3GPP compliant. Moreover, the proposed BDE does not require knowledge on how the MAC scheduler operates which enhances its interoperability. Experimental results show that the BDE satisfies QoS requirements involving per-user average or tail packet delay metrics with significantly less bandwidth than static approaches and baseline online learning schemes.

Overall, our paper is structured as follows. In Sec. \ref{rel}, we describe several related works and compare them to ours. In Sec. \ref{sysa}, we present the system architecture. In Sec. \ref{pf}, we formulate the overall objective of the BDE as a DP problem. In Sec. \ref{sola}, we describe the solution approach based on RL. In Sec. \ref{impl}, we describe the implementation of the proposed BDE on the Amarisoft testbed. In Sec. \ref{exp}, we conduct experiments to evaluate the proposed BDE. Section \ref{concl} concludes the paper.

\section{Related Literature}
\label{rel}
A simlar concept to the bandwidth demand is the effective bandwidth which has been studied extensively several decades ago \cite{effective2,effective}. The works in this area showed that to satisfy constraints on the tail of buffer occupancy distribution, we may view a traffic source as if its bitrate is constant when the service bitrate of the system is also constant. This constant bitrate of the traffic source was called effective bandwidth and it has been used for buffer sizing and admission control. Under certain conditions, the effective bandwidth can be calculated numerically, e.g., $M/G/1$ queueing systems. 

Unfortunately, these conditions do not typically hold in wireless, where the service rate is random and time-varying. Moreover, the RAN is better modeled as a multiserver queueing system where each PRB corresponds to a server. In addition, the complicated MAC service disciplines in LTE/5G do not facilitate the use of effective bandwidth theory. Also, the aforementioned results are typically asymptotic and may not be appropriate for bandwidth adaptation. Nonetheless, we believe that incorporating effective bandwidth theory in a data-driven method such as the one proposed here is a promising approach. However, it is not clear how this can be done and we were not able to do so.

More recently, the interest in QoS guarantees in queueing systems has been resurfacing due to the advances in model-free learning methods \cite{smallone,smalltwo}. In \cite{queuelearn}, the authors propose an RL approach to provide probabilistic guarantees for end-to-end packet delays with minimum service rates at the traversed nodes. To do so, the reward in each timeslot is composed by a term that counts the packets received within the delay threshold and by a term that penalizes the sum of the service rates, i.e., the actions. In \cite{learnadm}, the same authors propose an RL formulation for admission control that provides end-to-end delay guarantees with minimum rejections.

Although we also consider packet delay metrics for the development of the BDE, we do not have direct control of the service rate of the queueing system. In our case, the BDE controls the PRBs allocated to the MAC scheduler which in turn affects the bitrate delivered to the NS. Note that this bitrate is also affected by the wireless channel conditions which are hard to model. Overall, the problem that the BDE needs to solve involves the dynamic dimensioning of a queueing system. Thus, it is not clear how to apply many of the existing RL approaches on queueing systems \cite{liubai, MQNlearn, queueinfocom} where the objective is to learn a good scheduling policy.

The authors in \cite{veciana} use MABs to learn how a fixed amount of bandwidth should be split among different NSs to maximize the overall system utility, which involves packet delay metrics. The authors address the coupling between current actions and future rewards by applying actions whenever the queue empties. If the queue does not empty for a long period of time, all the packets in the queue are dropped. The analysis of their algorithm is based on the prior work in \cite{interrupt}. 

Similarly, the authors in \cite{distutil} also develop an algorithm that learns how to split the resources to maximize the total network utility. However, as mentioned earlier, utility maximization does not necessarily imply satisfaction of the SLAs. Moreover, such algorithms do not scale well as the number of NSs increases since the state-action space explodes.

In \cite{zussman}, the authors dynamically allocate bandwidth to a NS based on a metric called REVA. This metric computes the amount of PRBs that very active bearers can obtain. These are bearers that attempt to obtain more than their fair share of PRBs, e.g., file transfer bearers. By combining REVA with the spectral efficiency of each user, the authors compute the bitrate of each bearer and then find the PRBs that the NS demands. Thus, the allocation scheme in \cite{zussman} relates to a BDE when the desired QoS involves bitrates and the NS includes various types of bearers. The authors develop a model to predict REVA and the spectral efficiency of the NS. Then, they use these estimates to find the PRBs that the NS needs in the near future. The framework in \cite{zussman} is interesting, however it is not clear how it can be extended for QoS involving packet delay metrics. 

In general, traffic prediction in mobile networks has received a lot of interest in recent years \cite{microscope, vicenzo, howtoslice, tontraffic, spatiotempmodel, metropolis}. However, the objective of the network function developed here is not to predict the future traffic of the NS, e.g., the future number of traffic flows. Instead, the BDE wishes to learn the relation between the current traffic and the required allocated resources.

In \cite{ATHENA}, the authors implement a MAC scheduler based on contextual MABs in the srsRAN software solution. Here, we instead develop an algorithm that adapts the available PRBs used by the MAC scheduler which in our case is the default proportionally fair scheduler of the AMARI Callbox.

The authors in \cite{orion} present Orion, a RAN slicing system that allows the use of resource sharing algorithms in the BS to honor the SLAs of NSs without wasting PRBs. The authors implement Orion in OpenAirInterface (OAI), an open-sourced LTE platform. Here, we instead implement our BDE on the Amarisoft testbed which is closed-source. However, it allows long system-level LTE emulations with multiple users.


\section{The System Architecture}
\label{sysa}
In this section, we present the considered system architecture which was originally introduced in \cite{isolation}. The overall goal of this architecture is to satisfy the SLAs of all NSs with minimum provisioned bandwidth at the BS via dynamic resource adaptation. The SLAs considered here involve statements of the form "the packet delay metric $Q_i(t)$ of slot $t$ of NS $i$ should be less than $Q_c$ for at least $P_i$ fraction of slots". For example, the SLA of NS $i$ may state that for $P_i=95\%$ of timeslots, the average packet delay $Q_i(t)$ in the slot must be less than $Q_c=50$ ms. To this end, we consider two network functions; the Bandwidth Demand Estimator (BDE) and the Network Slice Multiplexer (NSM). 

The goal of the BDE is to perform dynamic bandwidth adaptation as explained earlier. At each slot $t$, the BDE observes the state of NS $i$ denoted by $X_i(t)$, and estimates the number of PRBs $W_i(t)$ that needs to be allocated to the MAC scheduler so that $Q_i(t) \leq Q_c$. The bandwidth $W_i(t)$ is the bandwidth demand of NS $i$ at slot $t$.

The NSM receives as input the bandwidth demand $W_i(t)$ of each NS $i$ at each slot $t$. Then, it decides which NSs receive their requested $W_i(t)$ PRBs at their MAC schedulers. The NSM makes this binary decision $u_i(t)$ for each NS $i$ by considering the limited bandwidth at the BS, any priorities between the NSs, and that the QoS of each NS $i$ needs to be met at least for $P_i$ fraction of slots.

\begin{figure}
\centering
\includegraphics[width=0.9\linewidth]{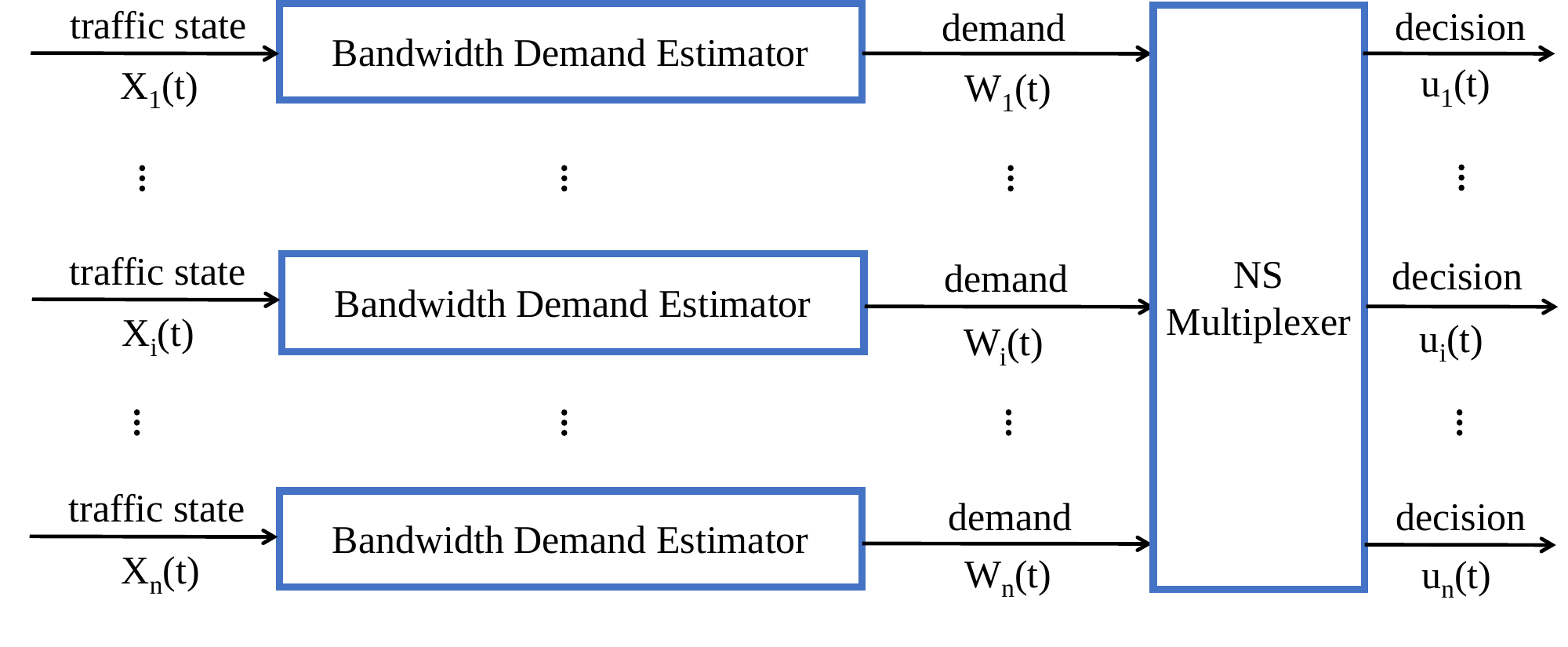}
\caption{First, the BDE observes the state $X_i(t)$ of NS $i$ and estimates the PRBs $W_i(t)$ needed to deliver the desired QoS at slot $t$. Second, the NSM receives all bandwidth demands $\{W_i(t)\}_{i \leq n}$ and decides which ones to satisfy given the limited bandwidth at the BS and the SLAs of all NSs.}
\label{system}
\end{figure}

The final step is to provision enough PRBs at the BS s.t. the NSM can accept the demands $W_i(t)$ of each NS $i$ for $P_i$ fraction of time. As mention earlier, this typically involves computing a percentile of the random process of total bandwidth demand. Once the amount of provisioned PRBs is found, the costs of the NSs are computed and the SLAs are formed. Overall, the two functions enable the NO to satisfy the SLAs of many NSs with reduced PRBs at the BS via statistical multiplexing. Figure \ref{system} depicts the system architecture.

Notice that the architecture enables learning to happen separately for each slice. Also, the learning task and the task of resolving resource contention is performed separately by the BDE and the NSM respectively. Although the complexity of the NSM depends on the number of NSs, once the bandwidth demands are known, the NSM needs to solve only a simple binary knapsack problem. Hence, the architecture enables a resource allocation approach that scales better than schemes that learn online how to maximize the total network utility.

A NSM that requires the least provisioned bandwidth at the BS for markovian demands and considers performance isolation constraints was proposed in \cite{isolation}. Any further modifications regarding the NSM and its resource provisioning mechanisms are out of the scope of this paper. Here, we focus only on developing a BDE for RANs.

The design of the BDE depends greatly on QoS requirements. If the desired QoS is to deliver a bitrate $B_c$ to each user of the NS, the BDE can compute $W_i(t)$ based on the current MCS of each user which composes the state $X_i(t)$. However, for packet delay requirements, the design is much more challenging, especially in RANs. In the remainder of this paper, we develop a BDE for packet delay requirements. 

\section{Problem Formulation}
\label{pf}
\subsection{Motivation for a dynamic programming formulation}
The main objective of the BDE is to adapt the allocated PRBs to the MAC scheduler of the NS so that the QoS is met with minimal bandwidth over time. To do so, at the beginning of each slot $t$, the BDE observes the state of the NS $X(t)$, which includes any information that helps the BDE to estimate the amount of bandwidth $W(t)$ needed to meet the QoS. 

\begin{figure}
\centering
\includegraphics[width=\linewidth]{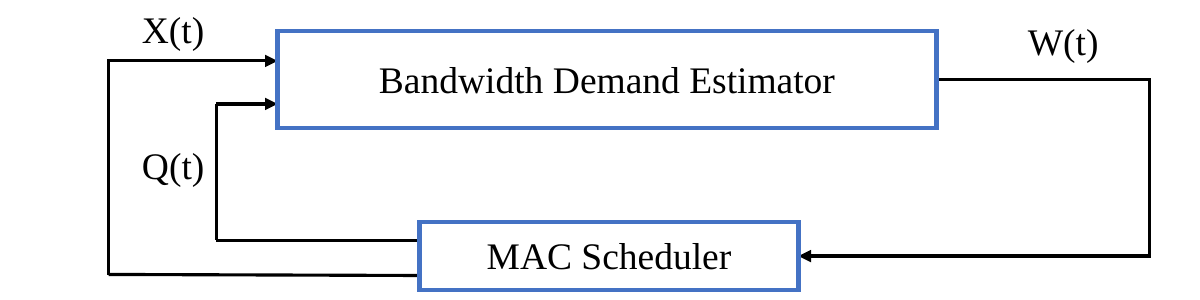}
\caption{At the start of slot $t$, the BDE observes the state $X(t)$ of the NS. Then, it estimates the number of PRBs $W(t)$ required to deliver the desired QoS throughout slot $t$ and allocates it to the MAC scheduler of the NS. At the end of slot $t$, it computes the cost $c(t)$ based on $W(t)$, the QoS feedback $Q(t)$ and the desired bound $Q_c$.}
\label{BDEfig}
\end{figure}

Here, we primarily focus on QoS involving packet delay requirements. We wish that a packet delay metric of interest $Q(t)$ regarding the packets delivered within $[t,t+D)$ to be smaller than $Q_c$, where $Q_c$ is a value defined in the SLA between the customer and the NO. At the end of each slot $t$, the BDE receives QoS feedback $Q(t)$ to fully evaluate the cost $c(t)$ of its action $W(t)$ as shown in Fig. \ref{BDEfig}.

Notice that small bandwidths $W(t)$ may satisfy the QoS but may also lead to large queue lengths. As a result, an action that initially seems promising may transition the NS to a state where very large bandwidths are needed to meet the desired QoS. Thus, the current actions of the BDE affect the costs it receives in the future. For this reason, we design the BDE based on a DP problem formulation.

\subsection{State}
The state of the DP problem at time $t$ is the state of the NS $X(t)$. Here, we consider a three-dimensional NS state that contains the number of active users, the average of the MCSs over all the users, and the number of bits awaiting transmission at the MAC scheduler at time $t$. Also, we discretize the state space by aggregating similar values of each component to obtain a finite state space $\mathcal{X}$.

Notice that to capture the effect of the current actions on the future, one component of the state needs to describe the queue length. Also, at least one component needs to describe the incoming traffic. In our experiments, all users within a NS produce packets with the same length and frequency, thus the number of active users is sufficient. Next, the service rate of the system is summarized by the average MCS over all users which is affected by the channel conditions. We note that the MAC scheduler selects the MCS for each user by considering their Signal to Interference plus Noise Ratio (SINR) and a bound on the Block Error Rate (BLER). We describe the service rate of the system by the average MCS over all users instead of a vector of all the MCSs to maintain a low dimensional state. Overall, the state $X(t)$ needs to be low dimensional but also sufficiently describe the underlying queueing system at the MAC layer which is challenging.

\subsection{Action}
The action in the DP problem at slot $t$ is the bandwidth allocated by the BDE $W(t)$. Once again, we consider that the action space $\mathcal{W}$ is finite. Indeed, in LTE, we can allocate any number of PRBs in the set $\{1, 2, ..., 100\}$. However, we consider that the action space $\mathcal{W}$ is a subset of that set.

\subsection{Cost}
We model the cost $c(t)$ at slot $t$ as follows:
\begin{equation}
c(t) \: \eqdef \: W(t) + \lambda \mathds{1}_{Q(t) > Q_c}, \textrm{ where } \lambda > 0.
\label{cost}
\end{equation}

Notice that the first term penalizes large bandwidths. The second term is equal to the constant $\lambda$ if the QoS is violated, otherwise it is $0$. The above cost allows to consider the two learning tasks that the BDE needs to complete; learn the bandwidth needed to satisfy the desired QoS but also learn to allocate bandwidth efficiently. Given an action, the cost takes only two values which greatly reduces the size of the model. The selection of a value for parameter $\lambda$ is discussed later on.

\subsection{Transition probabilities}
Let $H^{t}$ be a vector that contains all the state-action pairs up to slot $t-1$. The next state $X(t+1)$ and the cost $c(t)$ are modeled as random variables that depend on the current and past state-action pairs as follows:
\begin{align}
& \prob(X(t+1) = x', c(t) = c| X(t)=x,W(t)=w, H^{t} = h) \nonumber \\
& = p(x', c|x,w), \: \forall t \in \mathds{N}.
\label{qprobs}
\end{align}

Therefore, given the current state-action pair, the next state and the cost follow the same joint distribution over time and do not depend on their previous values, i.e., the system satisfies the Markov property. This is a reasonable model given that state $X(t)$ contains information regarding the traffic, the channel conditions and the queues of the NS. 

For instance, to arrive at (\ref{qprobs}), we may consider that $X(t+1)$ and $c(t)$ are functions of $(X(t),W(t),n(t))$, where $n(t)$ is noise that models packets bursts and channel fluctuations within $[t,t+D)$ which clearly cannot be observed at time $t$ and be included at $X(t)$. The noise is drawn independently each slot from a stationary distribution with $\prob(n(t) = n|X(t)=x,W(t)=w) = g(n|x,w)$. The above are standard assumptions in infinite horizon DP problems \cite{bertsekas2}.


\subsection{Metric}
Next, notice that the BDE is a stochastic system that operates throughout the lifetime of the NS. Thus, it is natural to consider the minimization of the expected weighted sum of $\gamma^t c(t)$ over an infinitely long horizon. Thus, we consider the following metric which depends on the initial state $x_0$:
\begin{equation}
 J(x_0) \: \eqdef \: \E\left[\lim_{T_H \to \infty}\sum\limits_{t=0}^{T_H-1}\gamma^tc(t)\right] = \lim_{T_H \to \infty} \E\left[\sum\limits_{t=0}^{T_H-1} \! \! \gamma^tc(t)\right].
\label{metric}
\end{equation}

The quantity $\gamma \in (0,1)$ is a discount factor that guarantees the convergence of the infinite sum. It also reflects that costs in the far future matter less than immediate costs. The second equality follows from Fubini's Theorem for non-negative functions. Metric $J(x_0)$ is called the cost-to-go at state $x_0$. 

The BDE affects the above metric by its bandwidth allocations over time $\{W(t)\}_{t \in \mathds{N}}$. The BDE selects the allocated bandwidth $W(t)$ based on the current state $X(t)$ and on the history recorded so far $H^t$. Thus, the BDE draws $W(t)$ from a distribution $\prob(W(t)=w|X(t)=x, H^{t}=h)$ which defines a causal policy. In infinite horizon problems where the system transition probabilities satisfy (\ref{qprobs}), it suffices to consider stationary markovian policies when minimizing (\ref{metric}) as stated in \cite{bertsekas2}. Thus, it suffices to consider:
\begin{equation}
\prob(W(t) = w | X(t) = x, H^{t} = h) = \pi(w|x).
\label{policyprobs}
\end{equation}

\subsection{Optimization problem}
Given all the above, an optimal bandwidth allocation policy can be obtained by solving the following DP problem:
\begin{align}
\underset{\{\pi(w|x)\}}{\text{min.}} \qquad  & \E\left[\sum_{t=0}^{\infty}\gamma^tc(t)\right]  \nonumber \\
 \text{s.t.:} \qquad & \sum_{w \in \mathcal{W}}\pi(w|x) = 1, \: \forall x \in \mathcal{X}, \nonumber \\
 & \pi(w|x) \geq 0, \: \forall x \in \mathcal{X}, \forall w \in \mathcal{W}, \nonumber \\
  & X(0)=x_0.
\label{dynamicprob}
\end{align}

The first and second constraints require that $\{\pi(\cdot|x)\}_{x \in \mathcal{X}}$ are valid probability distributions. The third constraint indicates that we consider the optimization problem given the initial state of the system since we can observe it. Lastly, we consider $\gamma=0.99$ in our problem.

\subsection{Selection of cost parameter $\lambda$}
We conclude this section by providing insight on the choice of parameter $\lambda$ in the cost $c(t)$ in (\ref{cost}). To this end, observe that:
\begin{align}
 \E[c(t)| & X(t)=x,  W(t)=w] \nonumber \\
 & = w + \lambda\prob(Q(t) > Q_c | W(t)=w, X(t)=x) \nonumber \\
 & = w + \lambda p_v(x,w) \: \eqdef \: c(x,w).
\label{expcost}
\end{align}

The last equality follows from  assumption (\ref{qprobs}) which implies that the above probability does not explicitly depend on $t$. Therefore, the expected cost $c(t)$ given the state-action pair $(x,w)$ is the weighted sum of the bandwidth $w$ and the probability of QoS violation $p_v(x,w)$. Lastly, notice that it is reasonable to assume that $p_v(x,w)$ is decreasing w.r.t. $w$; for a fixed state, the more bandwidth we allocate, the less likely it is to violate the desired QoS. Next, we introduce the Q factor of a state-action pair $(x,w)$ under policy $\pi$:
\begin{align}
Q_{\pi}(x,w) \: & \eqdef \: \E[c(t) + \gamma J_{\pi}(X(t+1))|X(t)=x, W(t)=w] \nonumber \\
& = c(x,w) + \gamma j_{\pi}(x,w).
\label{Qfactors}
\end{align}

Once again, the stationarity of the second term follows from (\ref{qprobs}). The term $j_{\pi}(x,w)$ is the expected cost-to-go of the next slot when following policy $\pi$, given that the current state-action pair is $(x,w)$. Note that we may assume that $j_{\pi}(x,w)$ is decreasing w.r.t. $w$. This is justified since allocating more bandwidth will likely further reduce the packet queue of the NS, and thus transition the system to a more favorable state $X(t+1)$ with a smaller cost-to-go $J(X(t+1))$.

An important property of the Q factors is that they can be used to generate a policy $\pi'$ that is better than a given policy $\pi$. This is achieved by considering $\pi'$ such that for each state $x$, $\pi'(w_x|x)=1$, where $w_x = \arg\max_w Q_{\pi}(x,w)$. Indeed, several DP algorithms use this property to produce an improved policy at the end of each iteration \cite{sutton}.

Now, suppose that the customer of the NS imposes the following requirement. For a given state $x$, if applying bandwidth $w'$ reduces the probability of QoS violation by more than $\epsilon$ compared to the case where bandwidth $w$ is used, then bandwidth $w'$ should be preferred over bandwidth $w$. 

Such requirements force the network operator to choose a larger bandwidth $w'$ over $w$ if it significantly improves the QoS delivery, regardless of the bandwidth savings $w'-w$ obtained otherwise. Given all the above, we can mathematically state this requirement as follows:
\begin{align}
p_v(x, w') \leq p_v(x, w) - \alpha \Rightarrow Q_{\pi}(x,w') \leq Q_{\pi}(x,w) \nonumber \\
\forall w', w \in \mathcal{W}, \forall x \in \mathcal{X}, \forall \pi.
\label{req1}
\end{align}

Note that the above statement captures the preference of $w'$ over $w$. Indeed, during policy improvement, for each state $x$, the new policy considers the action $w_x = \arg\max_w Q_{\pi}(x,w)$. Thus, action $w'$ will be preferred over $w$.

Let $W_{\max}$ and $W_{\min}$ be the largest and smallest element of $\mathcal{W}$ respectively. By exploiting the aforementioned monotonicity of $p_v(x,w)$ and $j_{\pi}(x,w)$, it is easy to show that a sufficient condition for (\ref{req1}) is:
\begin{equation}
\lambda  \geq \frac{W_{\max} - W_{\min}}{\alpha}.
\label{cond1}
\end{equation}

Next, the customer of the NS may allow the network operator to choose bandwidth $w$ over a larger bandwidth $w'$ if the larger bandwidth $w'$ does not improve the QoS violation probability by more than $\epsilon$. If the customer does not give any such leeway to the network operator, then the network operator may always need to allocate $W_{\max}$ PRBs to the NS, and thus charge the customer higher prices. 

The mathematical statement of this requirement and its sufficient condition are identical to (\ref{req1}) and (\ref{cond1}) respectively, but with flipped inequality signs. Thus, both of the above requirements are satisfied when $\lambda  = (W_{\max} - W_{\min})/{\alpha}$. Specifically, we consider $\alpha=0.01$ in our algorithm.
\section{Solution Approach}
\label{sola}
\subsection{Proposed reinforcement learning algorithm}
Note that the objective function of (\ref{dynamicprob}) contains infinitely many terms. Even in finite horizon problems where $t \leq T_H$, the analysis of the expectation in (\ref{dynamicprob}) involves all the possible histories $h$ of size $t$ which amount to $\sum_{t=0}^{T_H}(|\mathcal{X}| \times |\mathcal{W}|)^t$. 

Due to their large size, DP problems are typically solved by exploiting Bellman's principle of optimality; the truncated optimal policy of a problem is also optimal for the corresponding tail subproblem \cite{bertsekas2}. Bellman's principle of optimality allows us to write recursive equations regarding the optimal cost-to-go $J^*(x)$ and Q factor $Q^*(x,w)$ for each state $x$ and action $w$. We can numerically solve these equations in infinite horizon problems using algorithms such as Policy Iteration (PI) or Value Iteration (VI) \cite[Chapter 4.3-4.4]{sutton}.

Both PI and VI are initialized by an arbitrary policy. In each iteration $k$ of PI, we perform policy evaluation and then policy improvement. In policy evaluation, we repeatedly update each estimated cost-to-go $J_{\pi_k}(x)$ until a stopping rule is met. In policy improvement, we compute each Q factor $Q_{\pi_k}(x,w)$  using the system's transition matrix in (\ref{qprobs}) and the cost-to-go $J_{\pi_k}(x)$. Then, we generate the next policy $\pi_{k+1}$ by considering $\pi_{k+1}(w_x|x)=1$, where $w_x = \arg\max_w Q_{\pi_k}(x,w)$ for each state $x$. This concludes the iteration $k$ of PI. The algorithm terminates when $\pi_{k+1} = \pi_k$ indicating convergence to the optimal policy. The VI algorithm operates similarly. It combines the policy evaluation and improvement within a single for-loop, and then 
generates the final policy as previously.

Note that both PI and VI are guaranteed to converge to an optimal policy. Moreover, the number of iterations required for convergence is typically small and thus the speed of the algorithm is also high. However, both the PI and VI require the system dynamics in (\ref{qprobs}) to compute the Q factors. Unfortunately, the system dynamics are often not known in practice which is also the case for our problem.

Due to the above difficulties, we consider an RL approach to approximately solve the DP problem in (\ref{qprobs}) in a data-driven manner. There are multiple RL algorithms available in the literature \cite{sutton}. The majority of them periodically approximate the Q factors $Q_{\pi}(x,w)$ and the cost-to-go values $J_{\pi}(x)$ online using the currently available data $\{(X(\tau), W(\tau), c(\tau))\}_{\tau \leq t}$. In Monte Carlo methods, the data obtained online are used to directly approximate the Q factors and the cost-to-go values. In function approximation methods, parametric functions $Q_{\pi}(x,w;\theta)$ and $J_{\pi}(x;\theta)$ are considered and the data are used to update the parameters $\theta$. Once the Q factors are estimated, policy improvement is performed as described previously.

Here however, we consider a different approach. Instead of estimating the $Q_{\pi}(x,w)$ and $J_{\pi}(x)$ values, we periodically estimate the system dynamics $p(x',c|x,w)$ in (\ref{qprobs}) and then perform VI to obtain the optimal policy for the estimated system dynamics. We consider this approach primarily for two reasons. First, we experimented with Monte Carlo methods and they performed poorly. Second, we can leverage the monotonicity between the state, action, and cost to estimate the system dynamics $p(x',c|x,w)$ even for unseen $(x,w,x',c)$ tuples. Thus, the fast estimation of the system dynamics combined with the ability to quickly find the optimal policy through VI makes this approach particularly promising. In literature, such methods that plan over a learned model of the system dynamics are called model-based RL methods \cite{modelRL}.

Overall, the RL algorithm is structured as follows. Initially, we run an initial policy for a predefined number of slots $T_0$. Next, we loop through the following steps indefinitely. First, using the collected data and the monotonicity of the cost w.r.t. to the action, we estimate the transition probabilities of the system. Second, we perform VI to obtain an $\epsilon$-soft policy to ensure the exploration of the state-action space. Third, we apply the new policy for the next $T$ slots.

We note that the slot length $D$ is set to a small value to monitor the NS frequently. As a result, the number of users remains almost constant for multiple slots and so does the first component of the state, which aggregates similar numbers of users. Hence, we perform the above procedure separately for each value of the first component of the state, and then only consider the second and third state components for the transition matrix. This way we reduce the dimension of the transition matrix with little impact on performance. An overview of the RL algorithm is depicted in Algorithm \ref{ovalgo}.

\begin{algorithm}
 \For{$t \in \mathds{N}$}{
   get $X(t) = $ [users(t), avgMCS(t), bits-in-queue(t)]\;
   $x_1 = $ users(t)\;
   counts($x_1$) += 1\;
  \If{counts($x_1$) $\leq T_0$}{
  compute $W(t)$ using v-UCB1($x_1$)\;}
  \Else{compute $W(t)$ using $\epsilon$-soft-policy($x_1$)\;}
   wait $D$ seconds for the slot $t$ to end\;
   receive QoS feedback $Q(t)$ and compute $c(t)$\;
   log tuple $(X(t),W(t),c(t))$ to data($x_1$)\;
  \If{counts($x_1$) $\leq T_0$}{
   update the internal state of v-UCB1($x_1$)\;}
  \ElseIf{counts($x_1$)$\mod T == 0$}{
   estimate $p(x_2',x_3',c|x_2,x_3,w)$ from data($x_1$)\;
   perform VI \cite[Chapter 4.4]{sutton} to find Q factors\;
   update  $\epsilon$-soft-policy($x_1$)\;
   }
   }
 \caption{Overview of the proposed RL algorithm}
 \label{ovalgo}
\end{algorithm}

Lastly, we wish to mention that the proposed RL algorithm essentially performs receding horizon control with an infinite horizon \cite{receding}. To see this, note that VI provides the optimal policy for the infinite horizon DP problem given the current estimate of the transition probabilities. However, instead of solving the infinite horizon problem each slot given the updated estimate of the transition probabilities, we do so every $T$ slots. Ideally, we wish to update our policy every $T=1$ slots to immediately exploit the recently observed tuples $\{X(t),W(t),c(t)\}$. However, this may not be feasible due to the time complexity of VI and of the estimation of the transition probabilities.
\subsection{Monotonicity assumption}
In our algorithm, we consider the following type of monotonicity. If action $W(t)$ satisfied the QoS at time $t$, then it is natural to consider that any $w>W(t)$ would have also satisfied it at time $t$. Similarly, if action $W(t)$ did not satisfy the QoS at time $t$, then any $w<W(t)$ would also fail to satisfy it at time $t$. Similar statements can be made for some components of the state. For instance, the more the users are, the more the needed bandwidth is. However, we only consider the monotonicity of the cost w.r.t. the action for simplicity.

We leverage the aforementioned monotonicity in two parts of Algorithm \ref{ovalgo}. First, we use it to develop a variant of the Upper Confidence Bound 1 (UCB1) algorithm in \cite{orUCB} which we refer to as v-UCB1. Second, we use it to efficiently estimate the transition dynamics in line $15$.
\subsection{The v-UCB1 algorithm used for initialization}
We consider a MABs algorithm as our initial policy for the following reason. In queueing systems, low packet delays typically imply low packet queue lengths as suggested by Little's Law \cite{littlelaw}. Thus, state-action pairs $(x,w)$ that achieve low immediate costs may also achieve low Q factors under the optimal policy. As a result, by initializing Algorithm \ref{ovalgo} using MABs, we obtain multiple samples for promising state-action pairs $(x,w)$ which is useful for the estimation of the transition probabilities later on.

Specifically, we use the v-UCB1 algorithm for initialization which is identical to the standard UCB1 \cite{orUCB} but it also considers the following. If $Q(t) > Q_c$, then we consider that we observe the cost not only of $W(t)$ but of all smaller actions $w\leq W(t)$ as $w+\lambda$. Similarly, if $Q(t) \leq Q_c$, we consider that $\forall w\geq W(t)$, the cost is $w$. Thus, each slot we receive feedback for multiple actions $w$ not just for the selected action $W(t)$. We then use this feedback regarding the costs of multiple actions to update the internal state of the v-UCB1 algorithm accordingly. As a result, v-UCB1 avoids unnecessary exploration and reduces the regret.
\subsection{Estimation of the transition probabilities}
Next, we describe how the transition probabilities in (\ref{qprobs}) are estimated. Since we consider a different learning process for each number of users $x_1$ as shown in Algorithm \ref{ovalgo}, we only need to consider the average MCS over all users $x_2$ and the number of bits in the queue $x_3$ for the transition matrix. Thus, we create a matrix $P(x_1)$ that contains the number of times that each tuple $(x_2,x_3, w,x_2',x_3',c)$ occurred so far. Using the monotonicity assumption, we construct a new matrix $P''(x_1)$ as follows. For each occurrence of $(x_2,x_3,w,x_2',x_3',w+\lambda)$, we also consider  $\forall w' \leq w$, an occurrence of $(x_2,x_3,w',x_2',x_3'+1,w'+\lambda)$. Equivalently, if $w$ failed, then a smaller bandwidth $w'$ would also fail and result in a slightly worse queue length $x_3+1$. Similarly, we consider that if bandwidth $w$ succeeded, then a larger bandwidth $w'$ would also succeed and slightly improve the queue length. Next, we create a third matrix $P''(x_1)$ as a copy of $P(x_1)$. If a state-action pair $(x_2,x_3,w)$ has never occurred in $P''(x_1)$, we copy all of its corresponding elements from the second matrix $P'(x_1)$. Lastly, if a state $(x_2,x_3)$ never occurred in $P''(x_1)$, then we optimistically consider that $(x_2,x_3,W_{\min},x_2,0,W_{\min})$ occurred once. Equivalently, we consider that the minimum bandwidth satisfies the QoS with the least queue length possible. The transition probabilities $p(x_2,x_3,c|x_2',x_3',w)$ are then computed by the empirical pmf implied by matrix $P''(x_1)$. Algorithm \ref{estalgo} summarizes the above.

\begin{algorithm}
Create matrix P that contains the number of times each tuple $(x_2,x_3,w,x_2',x_3',c)$ occurred using data($x_1$)\;
$P' = P$\;
\For{each tuple $(x_2,x_3,w,x_2',x_3',c)$}{
\If{$c == w + \lambda$}{
\For{$w' \leq w$}{$P'(x_2,x_3,w',x_2',x_3' + 1,c)$ += $P(x_2,x_3,w,x_2',x_3',c)$}
}
\Else{
\For{$w' \geq w$}{$P'(x_2,x_3,w',x_2',x_3'-1,c)$ += $P(x_2,x_3,w',x_2',x_3',c)$}
}
}
$P'' = P$\;
\For{each tuple $(x_2,x_3,w,x_2',x_3',c)$}{
\If{$\sum_{x_2',x_3',c}P(x_2,x_3,w,x_2',x_3',c) == 0$}{$P''(x_2,x_3,w,x_2',x_3',c) = P'(x_2,x_3,w,x_2',x_3',c), \: \forall x_2',x_3',c$}
\If{$\sum_{w, x_2',x_3',c}P(x_2,x_3,w,x_2',x_3',c) == 0$}{$P''(x_2,x_3,W_{\min},x_2',0,W_{\min}) = 1$}
}
compute $p(x_2',x_3',c|x_2, x_3,w)$ from matrix $P''$
 \caption{Estimation of transition probabilities}
 \label{estalgo}
\end{algorithm}

\subsection{Construction of an $\epsilon$-soft policy}
The final part of the algorithm is to construct an $\epsilon$-soft policy using the Q factors obtained by VI. Although it is optimal to select the action $w$ with the largest Q factor $Q(x,w)$ w.p. $1$,  $\epsilon$-soft policies select $w$ w.p. $1-\epsilon + \epsilon/|\mathcal{W}|$ and all the others actions $w'$ w.p. $\epsilon / |\mathcal{W}|$. This way the RL algorithms can explore unknown state-action pairs and then evaluate them. Otherwise, the RL algorithm may get stuck at a poor policy. On the other hand, upon sufficient exploration, the $\epsilon$-soft policy is sub-optimal since the deterministic policy that selects the best action w.p. $1$ is better. In RL, this is known as the exploitation vs exploration tradeoff. Here, we set $\epsilon=0.01$.

\subsection{Discussion on parameter selection}
\label{discussion}
First, we discuss the choice of the slot length $D$. A small slot length allows the BDE to quickly adapt the bandwidth and exploit temporal variations that occur in a fast timescale. Also, it enables the BDE to receive QoS feedback frequently and take corrective actions fast so that the NS users do not experience poor QoS for long periods of time. It also speeds up the learning process since more data samples are observed within a fixed time interval. However, the time complexity of the algorithm and the presence of noise may forbid small slot lengths. In most of our experiments, we consider $D$ around $15$ seconds to provide fast adaption.

Next, the size of the state space $\mathcal{X}$ and the action space $\mathcal{W}$ greatly affect the performance, and the time and space complexity of the algorithm. As mentioned earlier, a large state-action space may increase performance since we can finely adapt the bandwidth based on minor changes at the state of the NS. However, the number of samples needed for the RL algorithm to converge increases and so does its computation and space complexity. Note that the VI and the estimation of transition probabilities in Algorithm \ref{estalgo} both have $\mathcal{O}(|\mathcal{X}_2|^2|\mathcal{X}_3|^2|\mathcal{W}|)$ time complexity. Thus, we consider sizes such as $|\mathcal{X}_1|=|\mathcal{X}_2|=|\mathcal{X}_3|=|\mathcal{W}|=3$ to include low, medium and high values for each state component and action.

The number of slots $T_0$ that the initial v-UCB1 algorithm is applied heavily depends on the size of the state-action space. In general, we wish that v-UCB1 stops iterating only after it has converged to the action with the highest expected reward for each state. Thus, at the very least, we wish that each state has the chance to be sampled once and set $T_0 = 100$.

Another crucial parameter is the duration of each episode $T$. The smaller $T$ is, the faster the policy is updated. Note that if $T=1$, then we perform VI and update our policy at the end of each slot. Hence, if $T=1$, the RL algorithm performs receding horizon control with an infinite horizon. Once again, we may not be able to do so due to the algorithm's time complexity. We consider $T=20$ in most of our experiments. 

\section{Implementation}
\label{impl}
\subsection{The Amarisoft testbed}
\begin{figure}
\centering
\includegraphics[height=5cm]{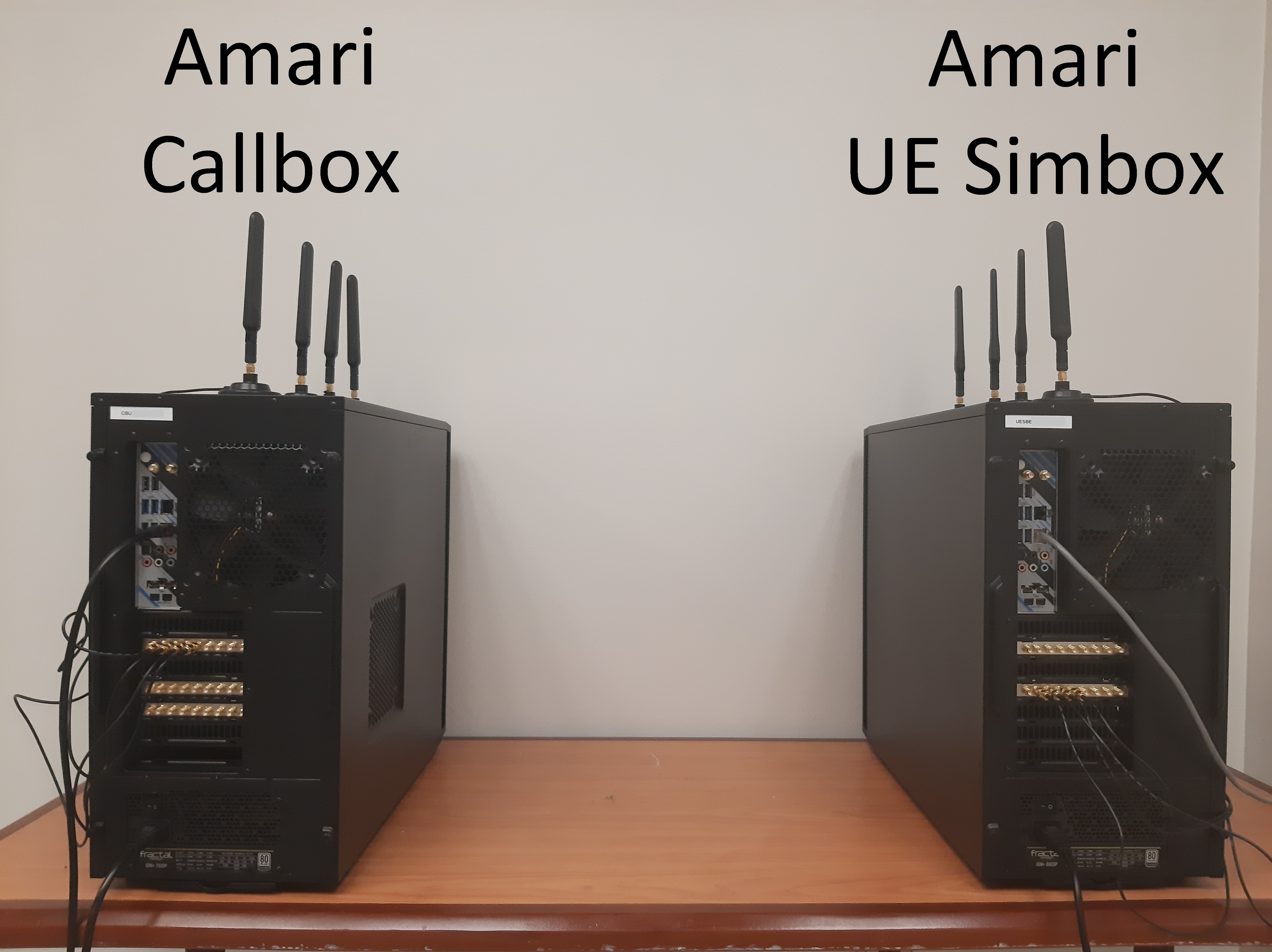}
\caption{The AMARI Callbox can emulate multiple LTE BSs and LTE CNs. Any commercial 3GPP compliant device, e.g., a smartphone, can connect to it via LTE once the Subscriber Identity Module (SIM) card is registered in the emulated LTE CN. The AMARI UE Simbox allows the emulation of tens of UEs without the need of new hardware devices. Both PCs are equipped with SDRs and antennas that allow them to communicate via LTE. The depicted setup uses 4 antennas on each PC which enables 2x2 MIMO.}
\label{Amarisoft}
\end{figure}

\begin{figure}
\centering
\includegraphics[height=4cm]{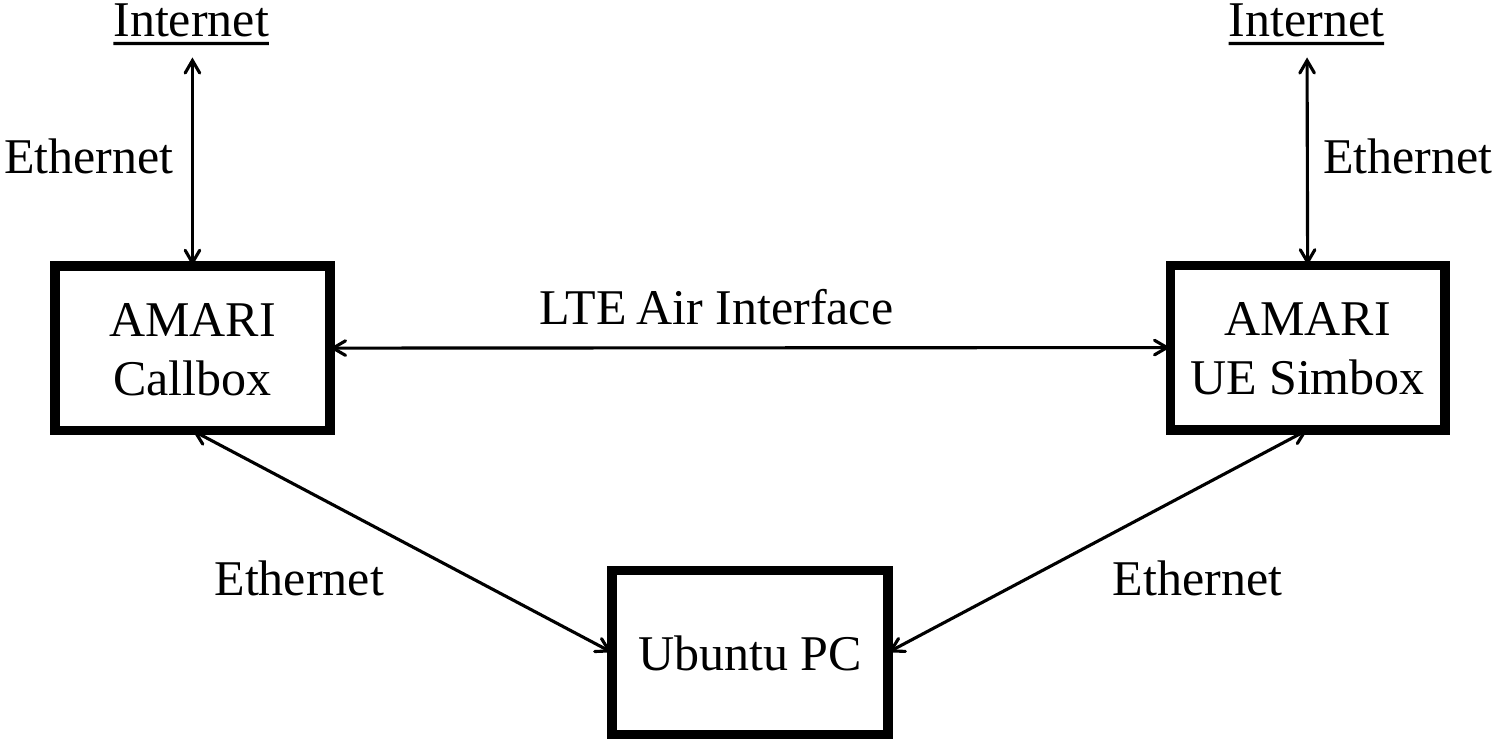}
\caption{The Amarisoft PCs communicate over-the-air via LTE. Each of them has two Ethernet ports. The first port connects to the Internet. The second port connects directly to an Ubuntu PC. The Ubuntu PC runs Algorithm \ref{ovalgo} and configures the Amarisoft PCs. It issues ping and API commands to the Amarisoft PCs, and downloads and parses their traffic logs online.}
\label{schematic}
\end{figure}

The testbed includes two integrated PCs; the AMARI Callbox Ultimate \cite{callbox} and the AMARI UE Simbox \cite{uesimbox}. The former can implement multiple BSs and a CN that connects them to the Internet. The latter can emulate up to 64 User Equipments (UEs). Both PCs include SDRs that allow them to communicate via over-the-air LTE or 5G. Each PC is connected via Ethernet to an Ubuntu PC where the BDE is implemented. Using this Ubuntu PC, we can issue ssh commands to the Amarisoft PCs to configure various parameters of the LTE cell, start test scenarios, download traffic logs and issue ping commands. The Amarisoft PCs are depicted in Fig. \ref{Amarisoft}. A schematic diagram of the testbed is shown in Fig. \ref{schematic}.

The two PCs can be used to perform system-level emulation of a cellular network with multiple UEs for many hours. This is an important feature since long emulation times are needed to emulate realistic traffic patterns in a NS. Due to this, we favored the Amarisoft testbed over other options, e.g., OAI, even though it is a closed-source commercial product.

\begin{figure}
\centering
\includegraphics[height=3cm]{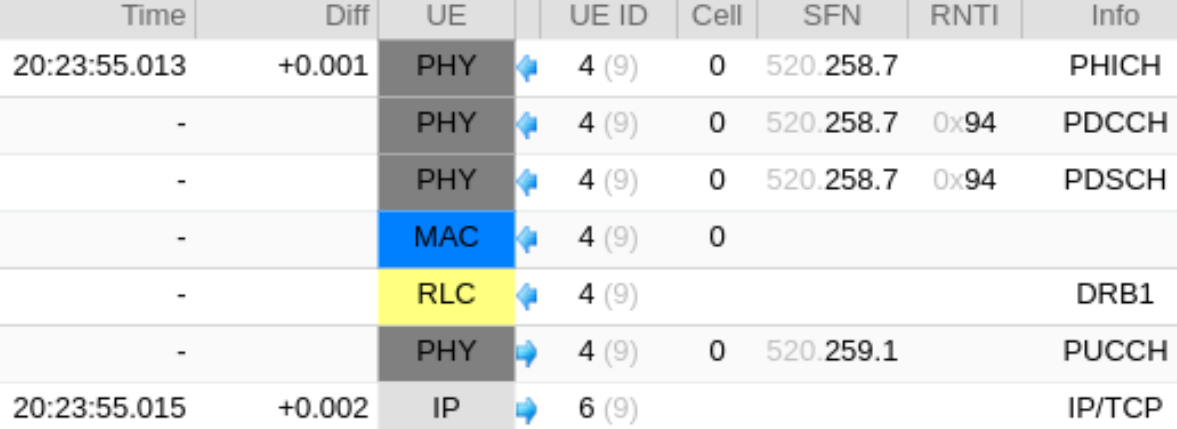}
\caption{A small excerpt of a traffic log. It shows the traffic in the UE Simbox across the LTE stack for two subframes, i.e., $2$ ms. The payload of the IP packets can be read as well. The BDE parses online such logs in their .txt format. The logs are similar to Wireshark logs enriched with LTE metadata.}
\label{WebGUI}
\end{figure}

A useful tool of the Amarisoft testbed is its WebGUI. It is an Apache server that visualizes logs with $1$ ms granularity containing the traffic of each PC in each layer of the LTE stack and various QoS metrics such as dropped packets, bitrates and Signal-to-Noise Ratios (SNRs). An example of a traffic log is depicted in Fig. \ref{WebGUI}. Such logs are used later on by our BDE.

The WebGUI may also be used to create and visualize emulation scenarios. The UE Simbox allows the creation of multiple UEs that may use different cellular protocols such as LTE and NB-IoT. Each UE may create a different type of traffic such as Real Transfer Protocol (RTP), Voice over IP (VoIP), and Hypertext Transfer Protocol (HTTP) traffic. We later consider that each such traffic type corresponds to a different NS. To create time-varying traffic patterns, we can set power on and power off events for each UE and set the duration of the application traffic during power on.

\subsection{Creating a test scenario}
The creation of a test scenario is done by configuring the parameters of a JavaScript Object Notation (JSON) file on the AMARI Callbox and the AMARI UE Simbox PCs. Regarding the LTE cell settings, we transmit on Band 7, which corresponds to center frequencies of $2680$ MHz and $2560$ MHz for DownLink (DL) and UpLink (UL) respectively. We deploy the maximum cell bandwidth of $90$ PRBs at the BS so that the BDE can try large bandwidths if needed. 

For simplicity, we use only one antenna for DL and one antenna for UL on each PC. We also enable channel simulation in the AMARI UE Simbox. Specifically, we consider the Gaussian channel model combined with a constant mobility model for the UEs. The number of UEs deployed depends on the experiment. The UE Simbox can emulate at most 64 UEs.

We create NSs with different types of traffic. The traffic types supported by the AMARI UE Simbox are RTP, VoIP, HTTP, and User Datagram Protocol (UDP). For each of these traffic types, various parameters can be configured. For instance, the VoIP traffic is parameterized by the packet size, the bitrate, the mean talking duration and the voice activity factor, i.e., the fraction of time that the UE generates traffic while in the Radio Resource Control (RRC) Connected state.

Thus, for each NS, we consider a JSON file that creates a number of UEs that all follow one of the previous traffic types. Next, to create time-varying traffic, we consider that each UE follows a sequence of on and off traffic periods. Hence, each UE follows a different sequence of on and off periods and the number of active users in the NS varies over time. 

\subsection{Simulating radio channel conditions}
As shown in Fig. \ref{Amarisoft}, the two PCs composing the mobile network are static and close to each other. Thus, the radio channel conditions are excellent which is not the case in real mobile networks. To simulate realistic radio channel conditions, the UE Simbox contains a radio channel simulator. In DL, the simulator modifies various measurements in the UEs to "fool" the LTE protocol into considering that the actual channel conditions follow some considered models. For instance, the channel simulator may use a constant speed mobility model for the UEs and the Extended Pedestrian A (EPA) channel model in \cite{ts36101}. Using these, it rewrites for each UE the Channel Quality Indicator (CQI), Reference Signal Received Power (RSRP) and Channel State Information (CSI). 

In UL, the simulator needs to know the actual pathloss between the UE Simbox and the Callbox. Then, it adjusts the actual transmission power of the UE Simbox so that the received signal power at the BS follows the considered channel models. To estimate the actual pathloss in UL, we first find the transmission power of the cell reference signal sent periodically by the Callbox and then compare it with the received power measured at the UE Simbox. In LTE, the first quantity is called the referenceSignalPower and is found in the System Information Block (SIB) broadcasted by the BS. In our case, it was equal to $9$ dBm. The second quantity is the Reference Signal Received Power (RSRP) which is measured by the UEs. In our case, it was $-84$ dBm for all emulated UEs. Thus, the DL pathloss is $93$ dB and we consider the same for the UL pathloss.

\subsection{Collecting the state of the network slice}
As mentioned earlier, the state of the NS is composed by three quantities; the number of active users, the average MCS over the users, and the total number of bytes awaiting transmission in the MAC layer. 

First, by active users, we consider UEs that are in the RRC Connected state in LTE, i.e, UEs that need to transmit data. We determine these UEs online by utilizing the remote Application Programming Interface (API) of the UE Simbox. In general, the number of active UEs is known to commercial BSs since information transfer and connection release messages are sent in the MAC layer whenever a UE needs to transmit data and becomes idle respectively.

Second, we also obtain online the MCS of each UE through the remote API. The MCSs provide information regarding the spectral efficiency of the UEs. They are known to any commercial BS since they are sent every 1 ms in both DL and UL directions. The average MCS over all active UEs is the second component of the traffic state.

Third, we obtain online the number of bytes awaiting UL transmission in each UE. We do so by scanning online the traffic logs of the UE Simbox for the most recent Buffer Status Reports (BSRs) \cite[Chapter 6.1.3.1]{ts36321}. The BSRs are MAC layer messages sent periodically by each UE to notify the MAC scheduler at the BS that there are bytes awaiting for transmission in the UE's buffer. Using the BSRs, the MAC scheduler avoids allocating more PRBs to each UE than the amount it needs to empty its buffer. Each BSR contains an index number that implies the number of bytes in the UE's buffer \cite[Table 6.1.3.1-1]{ts36321}. The third component of our traffic state is the total number of bytes in the buffers of all active UEs. Unfortunately, we are not able do find similar information for the DL buffers of a UE in the BS. Although the DL bytes awaiting for transmission are known to the BS and used by the MAC scheduler, we cannot observe them since the Amarisoft product is closed-source. Due to this, we primarily focus on the UL direction for our experiments. We note that obtaining and parsing traffic logs to find the most recent BSR takes around $0.5$ seconds.

\subsection{Adapting the bandwidth}
Once the traffic state is collected, the selected bandwidth is computed in Algorithm \ref{ovalgo}. To change online the PRBs available to the MAC scheduler of the NS, we use the remote API of the Callbox. It allows us to specify the number of PRBs that can be used for the Physical Downlink Shared Channel (PDSCH) and Physical Uplink Shared Channel (PUSCH) transmissions in DL and UL respectively. Thus, we can adapt the PRBs available to the NS both in DL and UL. 

\subsection{Computing the cost}
Once the bandwidth is changed, we determine its effect on the QoS metric $Q(t)$ after waiting $D$ seconds as stated in Algorithm \ref{ovalgo}. Throughout this time period, we issue ping commands at the BS to each active UE and log the ping responses. Once the $D$ seconds elapse, we compute the QoS metric $Q(t)$ based on the ping responses and check if the QoS is met by comparing it with $Q_c$. We note that ping commands measure round-trip delay but here we are interested in the one-way UL delay. For this reason, we need to estimate the DL delay and substitute it from the ping responses. To do so, we consider a preliminary scenario where only ping traffic is present. Then, we estimate the DL packet delay as one-half of the round-trip delay in the ping responses due to symmetry. Note that since we do not create any DL traffic in any of our test scenarios, this method approximates the actual DL delay in the test scenarios. The DL delay was $15$ ms. Alternatively, the One-way Active Measurement Protocol (OWAMP) \cite{owamp} can be used. Nonetheless, we follow our method to limit the installation of third-party software on our testbed. 

The code to implement the BDE on the Amarisoft testbed is available on GitHub\footnote{https://github.com/azulqarni/bandwidth-estimator-rl}. The code also includes a tool to create test scenarios. A setup such as the one in Fig. \ref{schematic} is required.

\section{Experimental Results}
\label{exp}
\subsection{Baseline schemes}
To properly evaluate the proposed RL algorithm, we compare its performance to the following three baseline schemes. 

\textbf{No adaptation:} This scheme does not perform any bandwidth adaptation. The no adaptation scheme simply considers that the state space is composed only  by a single possible state, i.e., $|\mathcal{X}|=1$. Therefore, it does not leverage any information regarding the current state of the NS. It simply uses the same instance of the v-UCB1 algorithm for the whole duration of the experiment. The no adaptation scheme acts as the reference point in our comparisons since currently deployed virtual networks receive a static amount of resources.

\textbf{v-UCB1:} The v-UCB1 scheme is identical to Algorithm \ref{ovalgo} for $T_0 = \infty$. Therefore, the v-UCB1 scheme creates a new instance of the v-UCB1 algorithm for every different value of the first state component as in Algorithm \ref{ovalgo}. However, it never performs VI. Thus, the v-UCB1 scheme is a myopic solution where we attempt to find the action with the lowest immediate cost regardless of its effect on the future. We compare the proposed algorithm with the v-UCB1 scheme to check whether RL is needed or if simpler myopic solutions suffice.

\textbf{MC control RL:} This scheme is identical to Algorithm \ref{ovalgo} but at the end of every $T$ slots, it performs the on-policy first-visit Monte Carlo control RL algorithm in \cite[Chapter 5.4]{sutton} instead of estimating the  system dynamics to do VI. Comparing the proposed algorithm with the MC control RL tests the benefit of model-based RL in our problem.

\subsection{Test scenario 1}
Here, we provide a simple scenario where our RL algorithm significantly outperforms the myopic v-UCB1 algorithm and explain why this happen. We consider that the NS consists of $10$ UDP users that are constantly transmitting at $1$ Mbps with packet lengths of $200$ bytes. The desired QoS is to ensure that average packet delay of each user is less than $Q_c=50$ ms. We set $T_0=T=20$ and $\epsilon = 0.1$ in the epsilon-soft policy used by the RL algorithm.

Note that the first component of the state $X_1(t)$ is constant over time $t$. Regarding the second component of the state $X_2(t)$ which aggregates similar values of the average MCS over all users, we consider the ranges $[0, 12)$ and $[12,28]$ which correspond to values $0$ and $1$ respectively. For the third component of the state $X_3(t)$ which aggregates similar values of the BSR index, we consider the ranges $[0, 20)$, $[20, 40)$, $[40, 61)$ and $[61, 62]$ that correspond to values $0$, $1$, $2$ and $3$ respectively. The available set of bandwidth allocations is $\mathcal{W} = \{24, 25, 90\}$ PRBs. We run this scenario for $500$ slots, i.e., for approximately $30$ minutes.

The v-UCB1 scheme quickly learns that the smallest bandwidth that satisfies the QoS is $24$ PRBs when the queue length is $X_3(t)=0$, $25$ PRBs when $X_3(t)=1$ and $90$ PRBs when $X_3(t)=2$. However, whenever $24$ PRBs are allocated to the NS, the queue length transitions to state $X_3(t)=2$. Then, the v-UCB1 algorithm allocates $90$ PRBs and the queue length transitions back to state $X_3(t)=0$. Thus, the v-UCB1 bandwidth allocations alternate between $24$ and $90$ PRBs. 

\begin{figure*}
\centering
\begin{minipage}{\linewidth}
\centering
Figures for test scenario 1
\vspace{1mm}
\end{minipage}
\begin{minipage}{0.45\textwidth}
\centering
\includegraphics[width=\linewidth]{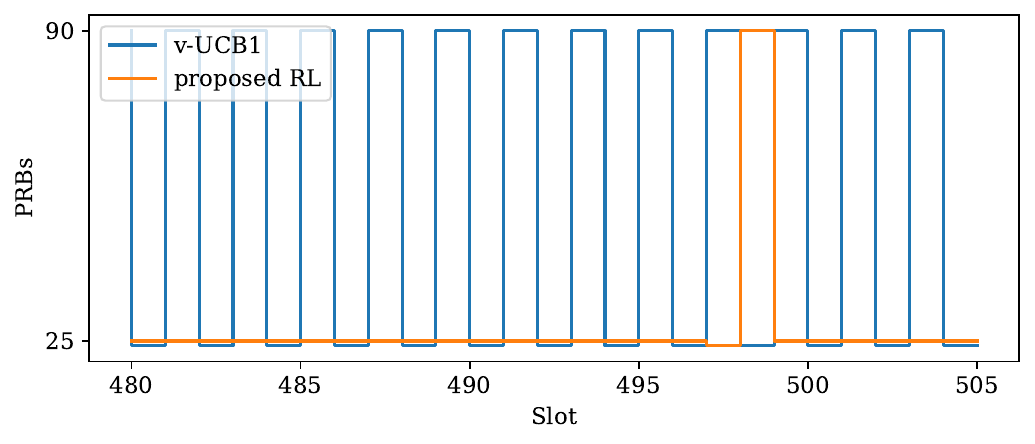}
\caption{The v-UCB1 scheme alternates between $24$ and $90$ PRBs, whereas the proposed RL algorithm learns to steadily allocate $25$ PRBs. At slot $497$, the RL algorithm allocates $24$ PRBs since an $\epsilon$-soft policy is used and sub-optimal actions are explored. As a result, the queue length increases and the RL algorithm uses $90$ PRBs at slot $498$ to reset it back to a small value.}
\label{test1_actions}
\end{minipage}%
\hfill
\begin{minipage}{0.48\textwidth}
\centering
\includegraphics[width=\linewidth]{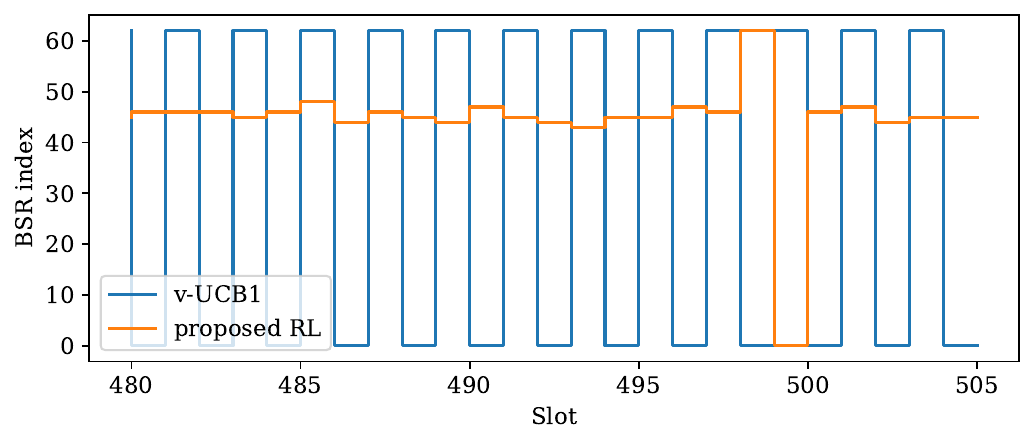}
\caption{The resulting queue lengths given the PRB allocations in Fig. \ref{test1_actions}. Due to its alternating PRB allocations, the queue lengths of the v-UCB1 scheme also alternate. On the other hand, the proposed RL algorithm stabilizes the queue length around a moderate value. The high and low queue lengths at slots $498$ and $499$ can be interpreted by observing Fig. \ref{test1_actions}.}
\label{test1_bsr}
\end{minipage}
\begin{minipage}{0.48\textwidth}
\centering
\includegraphics[width=\linewidth]{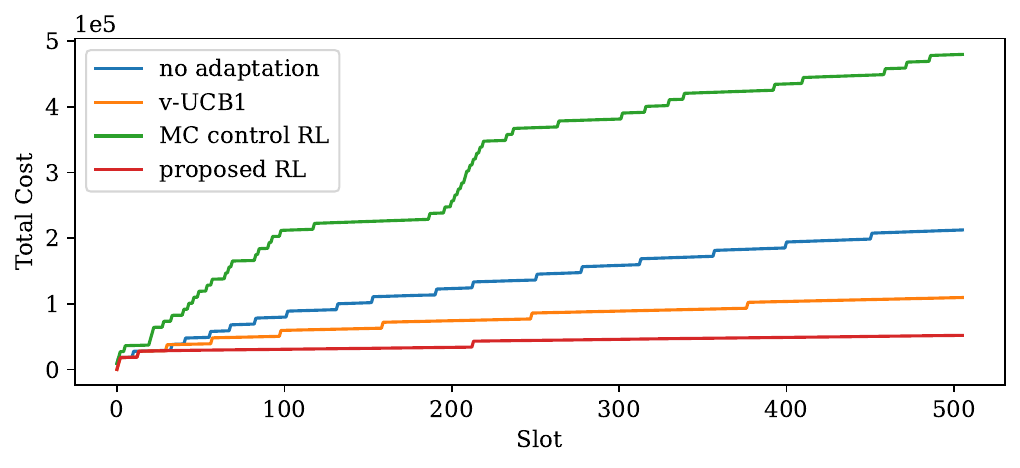}
\caption{The proposed RL algorithm accumulates the least cost over time. The cost at each slot depends on the allocated bandwidth and on the QoS deliver as shown in (\ref{cost}). It is worth-noting that the MC control RL scheme performed worse than the naive no adaptation scheme.}
\label{test1_cost}
\end{minipage}%
\hfill
\begin{minipage}{0.48\textwidth}
\centering
\includegraphics[width=\linewidth]{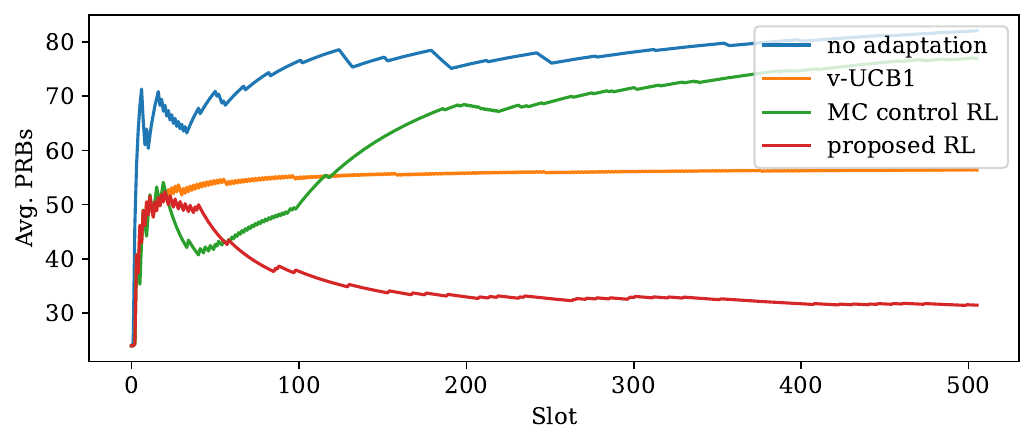}
\caption{The figure depicts the evolution of the average allocated bandwidth over time. Note that even in this test scenario where the number of users is constant, bandwidth adaptation significantly reduces the average allocated bandwidth. The RL algorithm achieves the largest bandwidth savings.}
\label{test1_avgbw}
\end{minipage}
\begin{minipage}{0.48\textwidth}
\centering
\includegraphics[width=\linewidth]{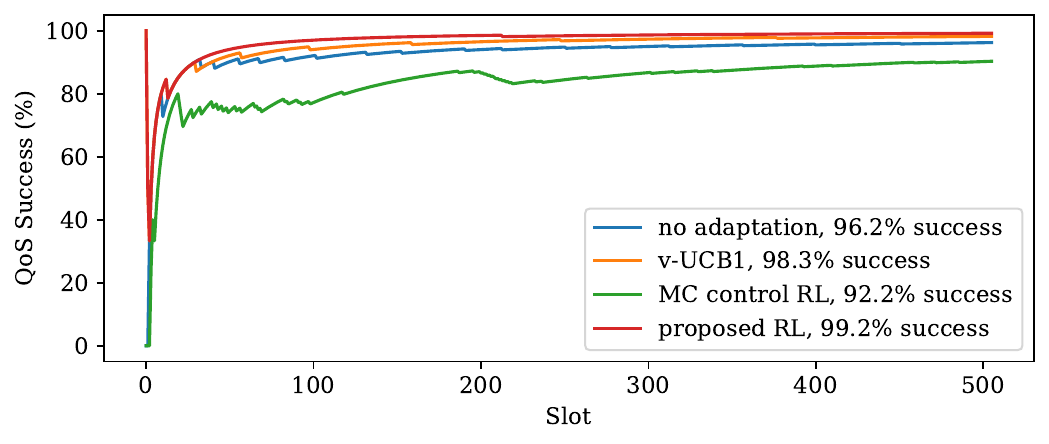}
\caption{The figure shows the evolution of the percentage of slots where the desired QoS is met. It follows from this figure and Fig. \ref{test1_avgbw} that the RL algorithm improves both the QoS delivery and the resource efficiency.}
\label{test1_qos}
\end{minipage}
\begin{minipage}{\linewidth}
\centering
\vspace{3 mm}
End of figures for test scenario 1
\end{minipage}
\end{figure*}

The optimal allocation policy is to allocate $25$ PRBs when $X_3(t)=0$ since an extra PRB is enough to maintain the queue length at $X_3(t)=0$ as observed from experimentation. Unfortunately, the v-UCB1 scheme is myopic and cannot consider such a policy. On the other hand, the RL algorithm in Algorithm \ref{ovalgo} is able to do so upon performing a single VI. As a result, the RL algorithm uses $25$ PRBs constantly instead of allocating $69$ PRBs on average as the v-UCB1 scheme does. The above are depicted in Fig. \ref{test1_actions} and Fig. \ref{test1_bsr}. We depict the full set of performance comparisons in Figs. \ref{test1_cost} to \ref{test1_qos}.

\subsection{Test scenario 2}
We create a more complex scenario where  the number of users varies over time. The scenario is run for $3500$ slots, where each slot's duration is $3.5$s, resulting in a total duration of $3.4$ hours. The users send UDP packets of $200$ bytes with $1$ Mbps bitrate. The users move with constant speed and path losses are updated based on their position.  The user ranges for the first state component are $[0,2)$, $[2, 8)$ and $[8, 12]$. The MCS and BSR ranges , and the QoS requirements are similar as before. The PRBs used are $\mathcal{W} = \{20, 40, 60, 90\}$.

\begin{figure*}
\centering
\begin{minipage}{\linewidth}
\centering
Figures for test scenario 2
\vspace{1mm}
\end{minipage}
\begin{minipage}{0.48\textwidth}
\centering
\includegraphics[width=\linewidth]{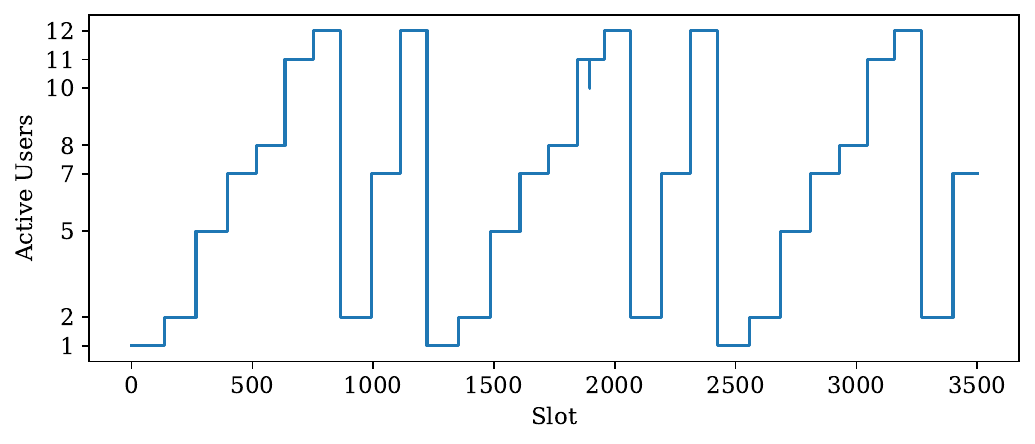}
\caption{The number of active users follows a periodic pattern that is terminated at slot $3500$. The total duration of the scenario is $3.4$ hours.}
\label{test2_users}
\end{minipage}
\hfill
\begin{minipage}{0.48\textwidth}
\centering
\includegraphics[width=\linewidth]{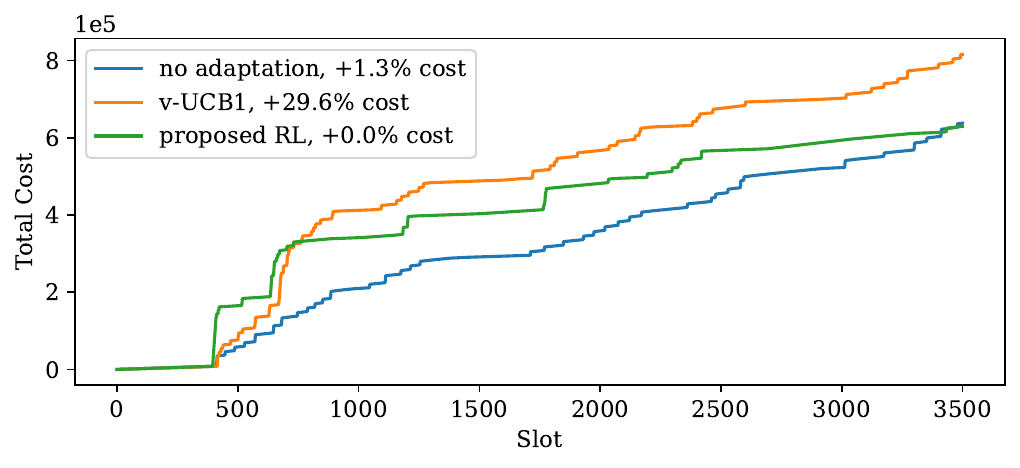}
\caption{The proposed RL algorithm updates the policy every $20$ slots and high costs occur when new users connect. However, the algorithm eventually learns to adapt the PRBs correctly and outperforms the other two schemes.}
\label{test2_cost}
\end{minipage}
\begin{minipage}{0.48\textwidth}
\centering
\includegraphics[width=\linewidth]{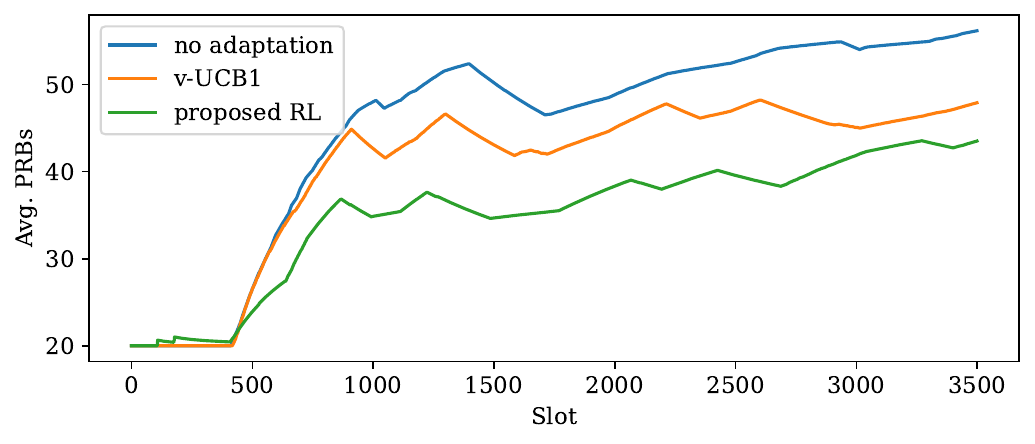}
\caption{The proposed RL algorithm significantly reduces the average PRB consumption compared to the no adaptation scheme.}
\label{test2_avgbw}
\end{minipage}
\hfill
\begin{minipage}{0.48\textwidth}
\centering
\includegraphics[width=\linewidth]{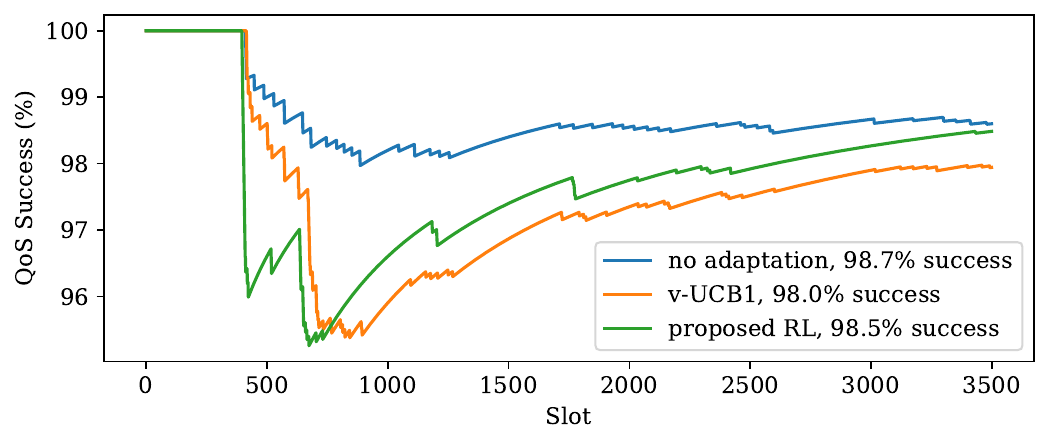}
\caption{The proposed RL algorithm initially delivers worse QoS than both other schemes. The figure shows that the algorithm's performance improves faster than the other two schemes.}
\label{test2_qos}
\end{minipage}
\begin{minipage}{0.48\textwidth}
\centering
\includegraphics[width=\linewidth]{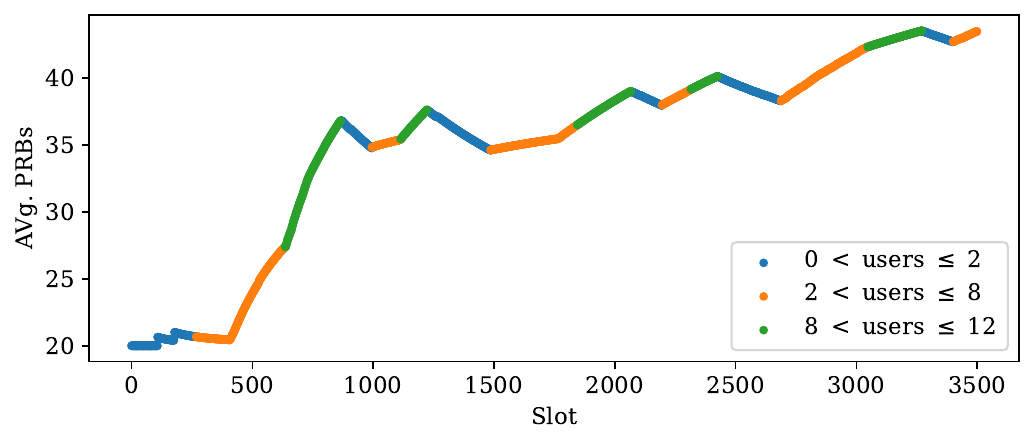}
\caption{The evolution of the average PRBs used by the proposed RL algorithm as the value of the first state component changes.}
\label{test2_BwRL}
\end{minipage}
\hfill
\begin{minipage}{0.48\textwidth}
\centering
\includegraphics[width=\linewidth]{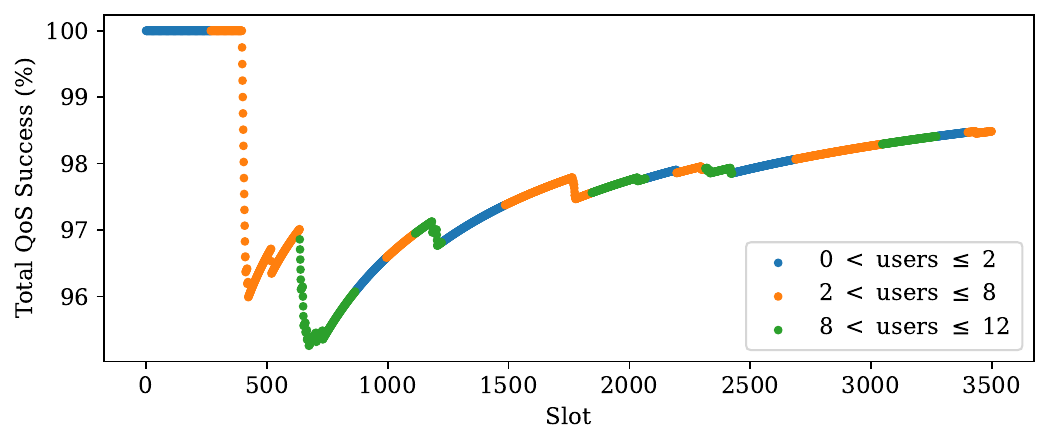}
\caption{The QoS success rate drops when the number of users changes. These drops diminish over time as the algorithm learns from past experience.}
\label{test2_QoSRL}
\end{minipage}
\begin{minipage}{0.48\textwidth}
\centering
\includegraphics[width=\linewidth]{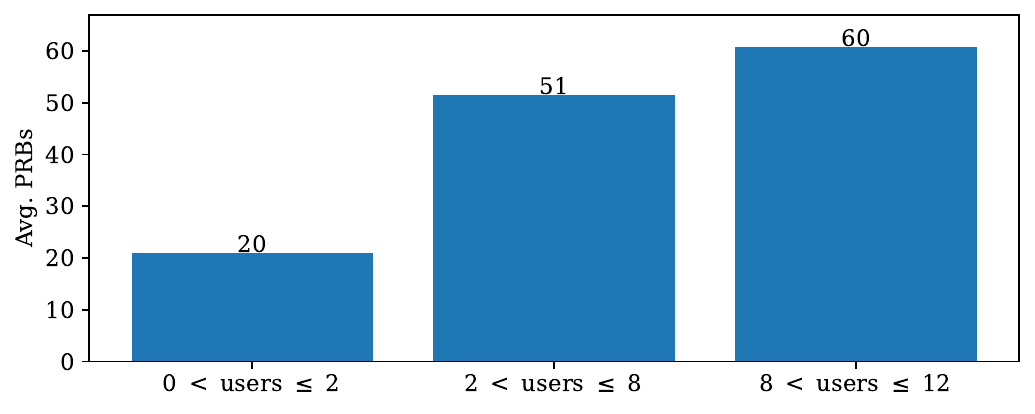}
\caption{The figure shows how the proposed RL algorithm adapts the PRBs. As expected, the average PRB usage increases as the users increase.}
\label{test2_perBwRL}
\end{minipage}
\hfill
\begin{minipage}{0.48\textwidth}
\centering
\includegraphics[width=\linewidth]{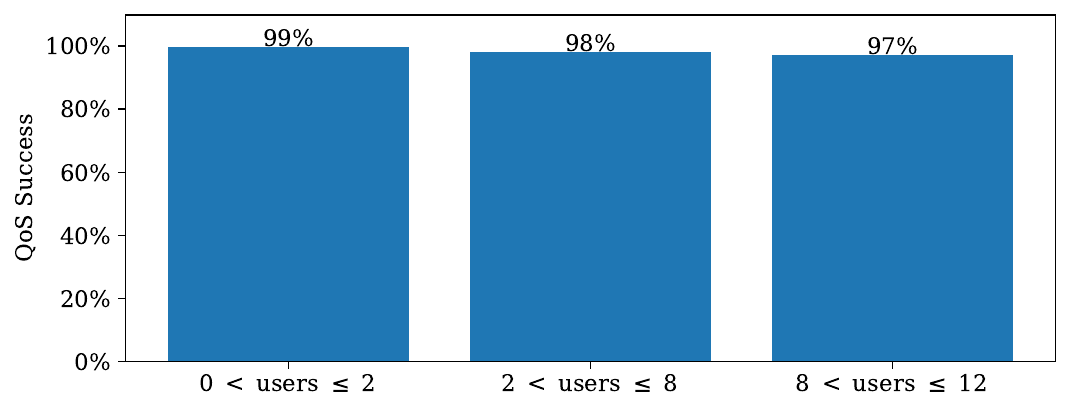}
\caption{The figure shows that the proposed RL algorithm delivers the desired QoS for a high fraction of slots for various numbers of active users.}
\label{test2_perQoSRL}
\end{minipage}
\begin{minipage}{\linewidth}
\centering
\vspace{3 mm}
End of figures for test scenario 2
\end{minipage}
\end{figure*}

We first plot the number of active users over time in Fig. \ref{test2_users}. Next, we depict the performance comparisons in Figures \ref{test2_cost} to \ref{test2_qos}. For illustration purposes, we did not plot the performance of the MC control RL scheme since it performed very poorly. Lastly, we provide more granular data for the performance of the proposed RL algorithm in Figures \ref{test2_BwRL} to \ref{test2_perQoSRL}.

\subsection{Test scenario 3}
We create a scenario where the number of active users varies over time. The total duration of the scenario is almost $3$ hours. The users transmit RTP packets of $200$ bytes. The mobility and channel model are the same as before. The state-action space $\mathcal{X} \times \mathcal{W}$ is similar as before with the exception of the first state component where we consider $\mathcal{X}_1= \{2,4,6\}$. 

We consider strict QoS requirements where we wish that $90\%$ of each user's packet to be lower than $Q_c=100$ ms. Hence, the desired QoS not only involves per user requirements but also tail packet delay metrics. This type of QoS is very difficult to analyze using Queueing Theory which showcases the value of data-driven approaches. 

\begin{figure*}
\centering
\begin{minipage}{\linewidth}
\centering
Figures for test scenario 3
\vspace{1mm}
\end{minipage}
\begin{minipage}{0.48\textwidth}
\centering
\includegraphics[width=\linewidth]{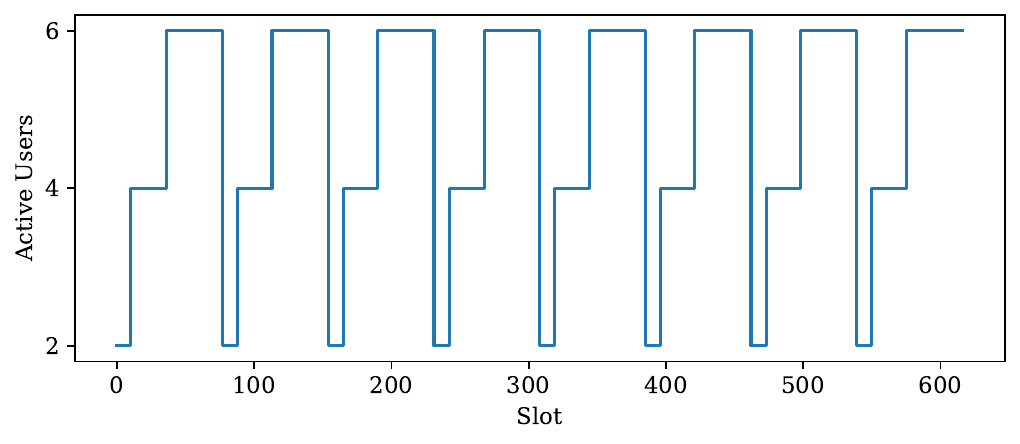}
\caption{Due to the strict QoS requirements, we consider lighter user traffic than previously. Otherwise, no bandwidth can deliver the desired QoS. The scenario's duration is almost $3$ hours and the slot length is $D=15$ seconds.}
\label{test3_users}
\end{minipage}
\hfill
\begin{minipage}{0.48\textwidth}
\centering
\includegraphics[width=\linewidth]{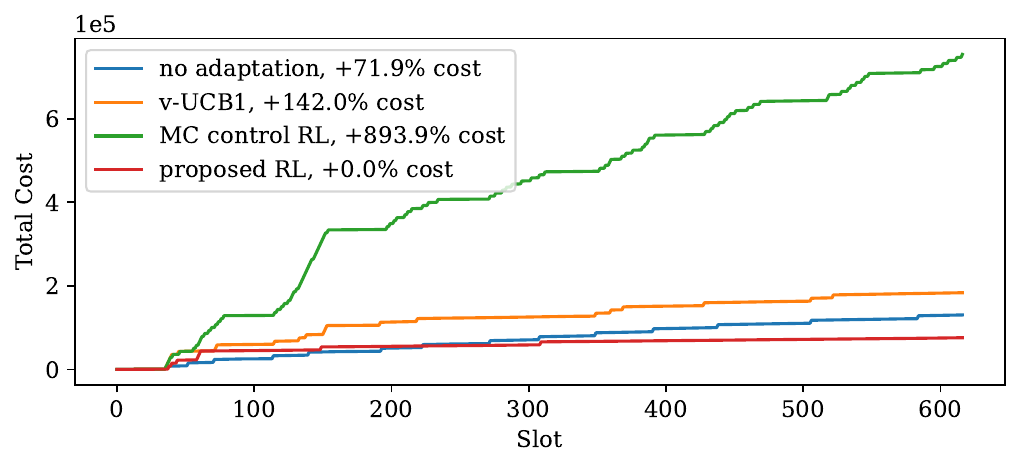}
\caption{The proposed RL algorithm initially performs worse than the no adaptation scheme due to its longer exploration phase. However, it eventually outperforms it. Once again, the MC control RL scheme performs poorly.}
\label{test3_cost}
\end{minipage}
\begin{minipage}{0.48\textwidth}
\centering
\includegraphics[width=\linewidth]{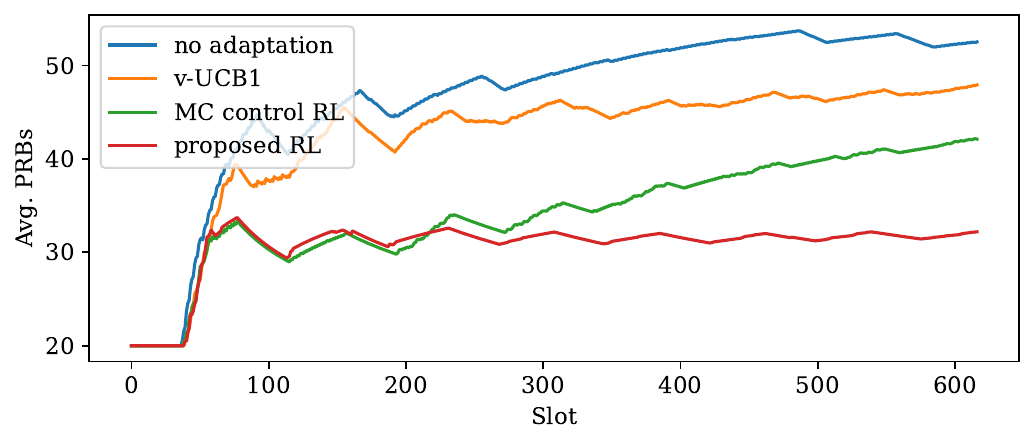}
\caption{Once again, the proposed RL algorithm significantly reduces the average PRB consumption compared to all other schemes.}
\label{test3_avgbw}
\end{minipage}
\hfill
\begin{minipage}{0.48\textwidth}
\centering
\includegraphics[width=\linewidth]{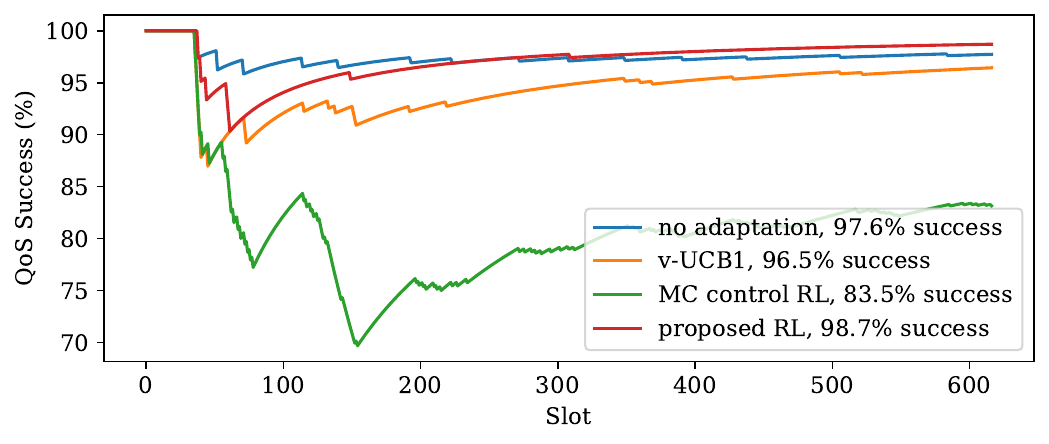}
\caption{The proposed RL algorithm also improves the overall QoS delivered to the NS. Thus, it outperforms the other schemes in both metrics of interest.}
\label{test3_qos}
\end{minipage}
\begin{minipage}{0.48\textwidth}
\centering
\includegraphics[width=\linewidth]{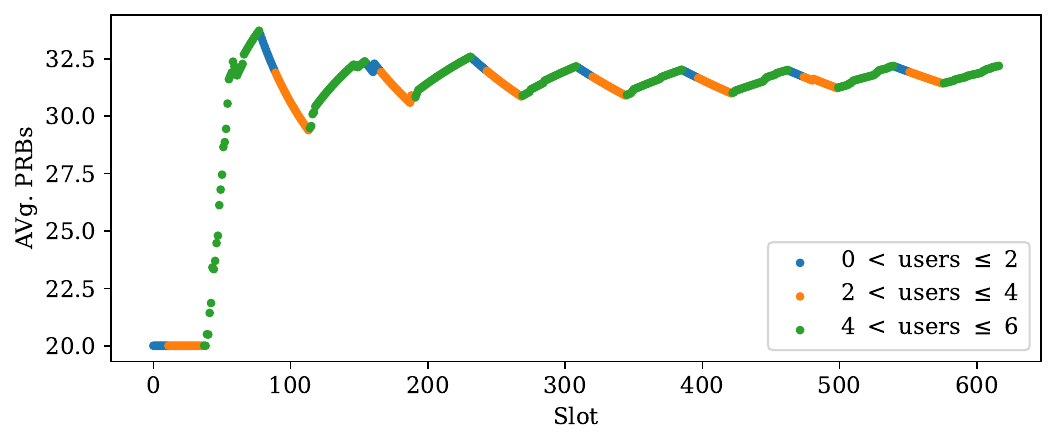}
\caption{The average number of PRBs used by the proposed RL algorithm converges to almost $32$ PRBs.}
\label{test3_BwRL}
\end{minipage}
\hfill
\begin{minipage}{0.48\textwidth}
\centering
\includegraphics[width=\linewidth]{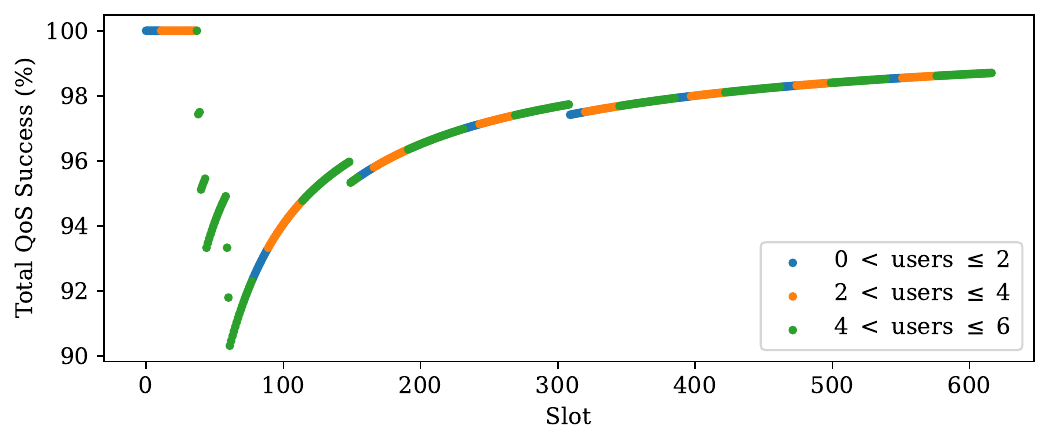}
\caption{The QoS success rate drops when the number of users changes. These drops diminish over time as the algorithm learns from past experience.}
\label{test3_QoSRL}
\end{minipage}
\begin{minipage}{0.48\textwidth}
\centering
\includegraphics[width=\linewidth]{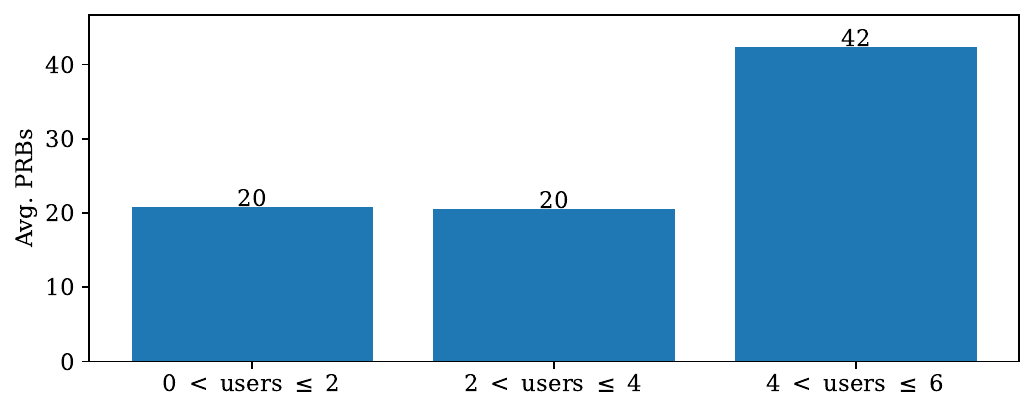}
\caption{Here, bandwidth adaptation is needed only when the users become $6$. Otherwise, the low value of $20$ PRBs is enough for both $2$ and $4$ users.}
\label{test3_perBwRL}
\end{minipage}
\hfill
\begin{minipage}{0.48\textwidth}
\centering
\includegraphics[width=\linewidth]{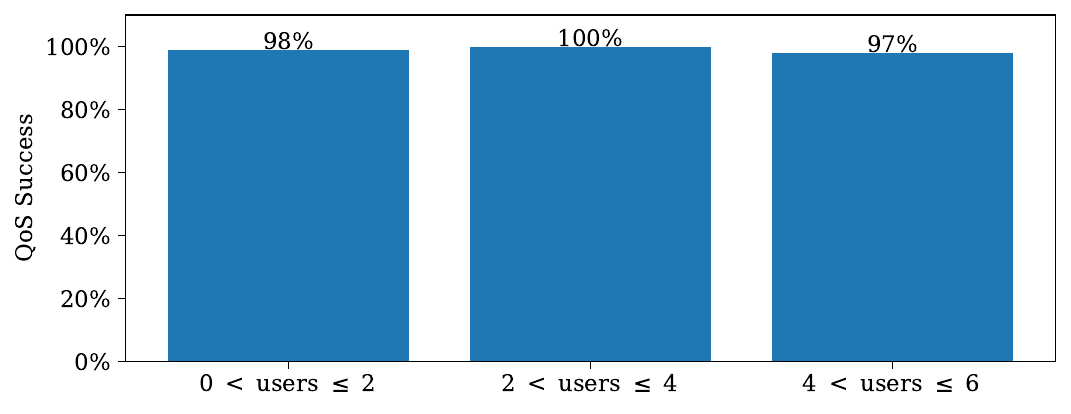}
\caption{The figure verifies that the proposed RL algorithm delivers the desired QoS for a high fraction of slots for all numbers of active users.}
\label{test3_perQoSRL}
\end{minipage}
\begin{minipage}{\linewidth}
\centering
\vspace{3 mm}
End of figures for test scenario 3
\end{minipage}
\end{figure*}

Moreover, we perform the estimation of the transition matrix and VI at the end of each slot, i.e., $T=1$ slots. Hence, the proposed RL algorithm performs standard receding horizon control with an infinite horizon. Due to the time complexity of VI, we increase the slot length to $D=15$ seconds so that the fraction of time where the BDE computes is $10\%$. We also reduce the number of slots that the initial MABs policy runs to $T_0=20$ slots. Regarding the MC control RL scheme, the number of slots per episode is kept to $T=20$ slots. The user activity diagram is depicted in Fig. \ref{test3_users}. The comparisons are in Figs. \ref{test3_cost} to \ref{test3_qos}. The efficiency of the proposed RL algorithm is shown in Figs. \ref{test3_BwRL} to \ref{test3_perQoSRL}.

\subsection{Execution time}
\begin{figure}
\centering
\includegraphics[width=0.85\linewidth]{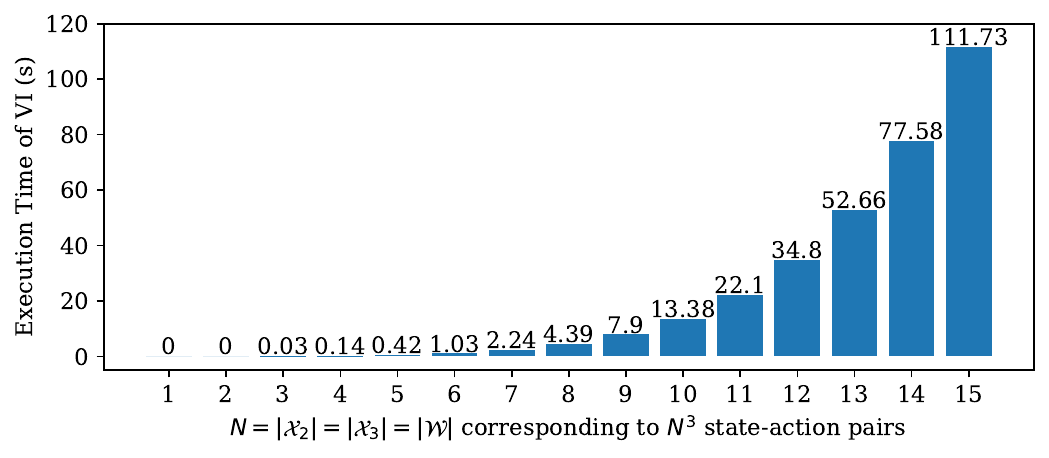}
\caption{We vary the size of the state-action in VI by considering each element takes $N$ value resulting in a space size of $N^3$. For each value of $N$, we compute the average execution time of VI after $10$ experiments. For $5^3=125$ state-action pairs, VI is completed within $0.5$ seconds.}
\label{timetest}
\end{figure}

Here we test the execution time of Algorithm \ref{ovalgo} and specifically its main bottleneck which is the VI performed as in \cite[Chapter 4.4]{sutton} with time complexity $\mathcal{O}(|\mathcal{X}_2|^2|\mathcal{X}_3|^2|\mathcal{W}|)$. In Fig. \ref{timetest}, we depict the execution times of the VI algorithm for various sizes of the state-action space. The total time required by all other operations is less than $1$ second.

\section{Conclusion}
\label{concl}
We implemented a BDE based on RL that adapts the PRBs of a NS based on its current traffic and the desired QoS. The experimental results on the cellular testbed show that the algorithm improves the QoS delivery and reduces the bandwidth when compared to some baselines. The incorporation of some domain knowledge enhanced the performance of the RL algorithm. In particular, the monotonicity between the the performance metric and the amount of allocated resources is common in many resource allocation problems. Here, we provided a simple method to exploit it via model-based RL. 

We note that the proposed BDE enables a system architecture that provides a scalable approach to satisfy the SLAs of multiple NSs. For future work, we wish to extend the current system architecture to address end-to-end QoS requirements spanning various network nodes.
\bibliography{references}
\bibliographystyle{IEEEtran}

\end{document}